\documentclass[a4paper,12pt]{article}
\pdfoutput=1
\usepackage[intlimits,centertags]{amsmath}
\usepackage{amssymb,amsfonts}
\usepackage{bbm}
\usepackage{mathrsfs}
\usepackage{cite}
\usepackage[scale={0.78,0.8}]{geometry}
\usepackage{multirow}
\usepackage{relsize}
\usepackage[subrefformat=parens,labelformat=parens]{subfig}
\usepackage{xcolor}
\usepackage{caption}
\usepackage{cancel}
\usepackage{graphics,graphicx}
\usepackage{float}
\numberwithin{equation}{section}

\def\beq{\begin{equation}}
\def\eeq{\end{equation}}
\def\beqa{\begin{eqnarray}}
\def\eeqa{\end{eqnarray}}
\def\a{\alpha}

\begin{document}

\pagestyle{empty}
\rightline{IFT-UAM/CSIC-16-127}
\rightline{MPP-2016-330}
\vspace{1.2cm}

\vskip 1.5cm

\begin{center}
\LARGE{Higgs-otic Inflation and Moduli Stabilization}
\\[13mm]
  \large{Sjoerd Bielleman$^{1}$,  Luis E. Ib\'a\~nez$^{1}$, Francisco G. Pedro$^{1}$, \\ Irene Valenzuela$^{2,3}$, and Clemens Wieck$^{1}$  \\[6mm]}
\small{
${}^1$  Departamento de F\'{\i}sica Te\'orica
and Instituto de F\'{\i}sica Te\'orica UAM/CSIC,\\[-0.3em]
Universidad Aut\'onoma de Madrid,
Cantoblanco, 28049 Madrid, Spain\\
${}^2$ Max-Planck-Institut fur Physik,
F\"ohringer Ring 6, 80805 Munich, Germany\\
${}^3$Institute for Theoretical Physics and
Center for Extreme Matter and Emergent Phenomena,\\
Utrecht University, Leuvenlaan 4, 3584 CE Utrecht, The Netherlands
\\[8mm]}
\end{center}
\begin{abstract}
We study closed-string moduli stabilization in Higgs-otic inflation in Type IIB orientifold backgrounds with fluxes. In this setup large-field inflation is driven by the vacuum energy of mobile D7-branes. Imaginary selfdual (ISD) three-form fluxes in the background source a $\mu$-term and the necessary monodromy for large field excursions while imaginary anti-selfdual (IASD) three-form fluxes are sourced by non-perturbative contributions to the superpotential necessary for moduli stabilization. We analyze K\"ahler moduli stabilization and backreaction on the inflaton potential in detail. Confirming results in the recent literature, we find that integrating out heavy K\"ahler moduli leads to a controlled flattening of the inflaton potential. We quantify the flux tuning necessary for stability even during large-field inflation. Moreover, we study the backreaction of supersymmetrically stabilized complex structure moduli and the axio-dilaton in the K\"ahler metric of the inflaton. Contrary to previous findings, this backreaction can be pushed far out in field space if a similar flux tuning as in the K\"ahler sector is possible. This allows for a trans-Planckian field range large enough to support inflation.\end{abstract}
\newpage

\setcounter{page}{1}
\pagestyle{plain}
\renewcommand{\thefootnote}{\arabic{footnote}}
\setcounter{footnote}{0}

\tableofcontents

%

\section{Introduction}

In the last few years, observational data from Planck and BICEP have sparked renewed interest in models of large-field inflation. Although evidence for cosmological tensor perturbations in the CMB is still elusive, present measurements allow for relatively large values of the tensor-to-scalar ratio $r$. Models of inflation with large $r$ typically point to trans-Planckian field excursions of the inflaton. In a trans-Planckian regime one can no longer ignore effects of quantum gravity, so that some ultraviolet completion of the theory is needed. String theory is the most promising candidate for such a completion, and it is thus natural to examine whether large-field inflation may be obtained in a controlled regime of string compactifications. Among the vast literature on large-field inflation models from string theory, cf.~\cite{Baumann:2014nda,Hebecker:2016dsw} for recent compilations of references, axion monodromy inflation \cite{Silverstein:2008sg,McAllister:2008hb} is a particularly intriguing idea. Here trans-Planckian field excursions may be naturally obtained, while the stability of the potential is protected by discrete shift symmetries inherent to string compactifications. 

The most obvious inflaton candidates in string theory come in two classes: First, closed-string moduli and the complex dilaton and second, open-string moduli. Explicit models with all required ingredients that yield sufficient $e$-folds of slow roll in agreement with observations, and at the same time feature a stable potential in all perpendicular directions, are notoriously difficult to construct. This is not surprising since every model of string inflation must address the problem of moduli stabilization. In models with open-string moduli the inflaton typically parametrizes the position of a D$p$-brane in the compact space, or the value of a continuous Wilson line. One of the examples considered recently involves position moduli of D7-branes in Type IIB orientifold compactifications \cite{Hebecker:2014eua,Arends:2014qca,Ibanez:2014kia,Ibanez:2014swa}. A particular minimal model  within this class is Higgs-otic inflation \cite{Ibanez:2014kia,Ibanez:2014swa}, in which the D7-brane position moduli correspond to the Higgs scalars of an MSSM-like string compactification. This is particularly attractive because it identifies the inflaton with a known particle and aspects like reheating follow quite naturally.\footnote{This is not what is usually called Higgs inflation, cf.~\cite{Bezrukov:2007ep}, which requires a nonminimal curvature coupling of the Higgs boson. In the Higgs-otic setup the Higgs couples minimally to gravity.} Although the model includes two neutral inflaton fields, $h$ and $H$, it was shown in \cite{Bielleman:2015lka} that isocurvature perturbations are exponentially supressed by the end of inflation, consistent with observations. 

In the present paper we extend the study of large-field D7-brane inflation models, with particular emphasis on the Higgs-otic benchmark. As of now, this model has been studied mostly from a local perspective based on the Dirac-Born-Infeld (DBI) and Chern-Simons (CS) actions of the D7-branes involved. ISD three-form fluxes in the background yield a monodromy and a mass term for the Higgs field, which implies a quadratic inflaton potential. However, this potential is flattened at large field values due to nontrivial kinetic terms arising from the DBI action. This leads to an approximately linear potential in the large-field regime, $V \sim  a \varphi$ \cite{Bielleman:2016grv}. Still, a fully realistic global model requires further ingredients. In particular, a consistent embedding of the D7-brane system requires a global compactification with stabilized moduli. Therefore, the aim of the present paper is to include the effects of moduli stabilization on the inflationary dynamics of Higgs-otic inflation. 

Moduli stabilization in Type IIB is best understood in terms of its low-energy $\mathcal N =1$ supergravity theory, cf.~\cite{Giddings:2001yu,Kachru:2003aw}. Thus, we need an $\mathcal N = 1$ description of the microscopic setup leading to Higgs-otic inflation, including sources which stabilize all moduli. As pointed out in \cite{Bielleman:2016grv}, one of the challenges is to reproduce the nontrivial kinetic term of the D7-brane from the DBI action. In that reference we have shown that, with only ISD three-form flux and in the global limit, the kinetic terms can be expressed in terms of a specific higher-derivative correction to the K\"ahler potential.\footnote{Cf.~\cite{Cecotti:1986jy,Cecotti:1986pk,Antoniadis:2007xc,Khoury:2010gb,Baumann:2011nm,Koehn:2012ar,Farakos:2012qu,Koehn:2012np,Koehn:2012te,Farakos:2013zsa,Gwyn:2014wna,Aoki:2015eba,Ciupke:2015msa,Broy:2015zba,Ciupke:2016agp,Cicoli:2016chb} for related studies of higher-derivative operators in supersymmetry and supergravity.} We argue here that, while the corresponding correction can be found and computed also in local supersymmetry, the identification of the correct operator is much more difficult once IASD fluxes are considered. The latter are necessary to describe non-perturbative contributions in the superpotential, which in turn are needed to stabilize K\"ahler moduli. However, even in this case the correct kinetic term can be found by expanding the DBI action---after including those additional fluxes---and expressing the result in terms of the correct supergravity variables.

In order to study moduli stabilization itself and its backreaction on the inflaton, we first consider a simplified KKLT-like setup in which the dilaton and complex structure moduli are stabilized supersymmetrically by fluxes and already integrated out. A single overall K\"ahler modulus is then stabilized by non-perturbative effects and an appropriate de Sitter uplift \cite{Kachru:2003aw}. It turns out that there are important backreaction effects which substantially modify the structure of the inflaton scalar potential once the K\"ahler modulus is stabilized and integrated out. Similar to our analysis here, the backreaction of stabilized moduli was studied in \cite{Buchmuller:2014vda} for supersymmetric stabilization, and in \cite{Buchmuller:2015oma,Dudas:2015lga} for the KKLT mechanism and other models in which the K\"ahler moduli break supersymmetry. Eventually, the modulus backreaction leads to an additional flattening of the effective potential, $V \sim a \varphi - b\varphi^2$. At the same time, the background fluxes must be chosen such that the mass of the inflaton is much smaller than that of the modulus. If this can be achieved, 60 or more $e$-folds of slow-roll inflation are possible. The tensor-to-scalar ratio then lies in the range $r\simeq 0.04-0.08$.

In addition to the backreaction induced by the K\"ahler moduli, there could be a backreaction from the complex structure and dilaton. In \cite{Baume:2016psm}, for example, such a backreaction feeding into the kinetic term of the inflaton was argued to severely limit the canonical field range, putting large-field inflation under pressure. However, as pointed out in \cite{Valenzuela:2016yny}, this problem seems to be particularly severe in models where the inflaton is a closed-string axion. Therefore, it is a natural question to ask whether the arguments of \cite{Baume:2016psm} apply when the inflaton is an open-string scalar. Based on simple toy models, we find that the constraints are far less severe in Higgs-otic inflation, as advanced in \cite{Valenzuela:2016yny}. Although a similar backreaction affects the kinetic term of the inflaton, the reduction of the canonical field range can be tuned by a flux choice, due to additional freedom not present in the models of \cite{Baume:2016psm}.

In summary, we present a model of large-field inflaton with full moduli stabilization. We find that the heavy moduli affect the simple structure of the Higgs-otic model studied in previous works, cf.~\cite{Ibanez:2014swa,Bielleman:2015lka,Bielleman:2016grv}, by inducing extra contributions to the scalar potential and kinetic terms. The most important effect arises from the backreaction of the K\"ahler moduli, which induces an additional flattening. This typically implies smaller values of the tensor-to-scalar ratio $r$. Although backreaction from complex structure moduli and dilaton may induce a reduction of the possible inflaton excursion, in the case of an open-string inflaton there is more freedom in the choice of flux parameters, allowing for trans-Planckian excursions large enough to obtain successful inflation.

The structure of the rest of this paper is as follows. In the next section we review and extend some of the general aspects of the action of D7-brane position moduli as present in the Higgs-otic system. In particular, we generalize the results discussed in \cite{Ibanez:2014swa} to include the DBI and CS actions in the presence of both ISD and IASD fluxes. We identify the correct  kinetic term of the inflaton in the presence of IASD fluxes, which leads to a flattening of the potential in the supergravity picture. Moreover, we comment on the possibility of describing this extended system in $\mathcal{N}=1$ supergravity with higher-order derivative operators. In Section 3 we combine the Higgs-otic open-string system with a single K\"ahler modulus in a KKLT-like setting. We study the associated backreaction on the inflaton potential following the analysis in \cite{Buchmuller:2015oma},
and show that, for an appropriate choice of flux parameters, consistent slow-roll inflation is achieved with a stable K\"ahler modulus. Moreover, we translate the nontrivial kinetic terms to the supergravity language and give numerical examples with the predicted CMB observables of the canonical inflaton variable. In Section 4 we investigate an additional backreaction induced by the complex structure moduli and the dilaton. Using flux stabilization as in \cite{Giddings:2001yu}, we show that the inflaton field range is almost unaffected when an appropriate mass hierarchy between the inflaton and moduli is achieved via a flux choice---just as in the K\"ahler sector. In Section 5 we review and discuss the different assumptions under which these results are obtained. In particular, we discuss the viability of the required flux tuning. We leave Section 6 for our conclusions. Finally, in Appendix A we illustrate that moduli-stabilizing fluxes may also yield $\mu$-terms for open-string moduli in a simple toroidal orientifold setting.

%

\section{Higgs-otic inflation in $\mathcal N = 1$ supergravity}
\label{sec:DBI}

Higgs-otic inflation \cite{Ibanez:2014kia,Ibanez:2014swa} is a recent proposal to realize large-field inflation in Type IIB string theory. We begin this section with a brief review of the model and the underlying compactification. We then extend the original setup by including IASD fluxes, in order to describe non-perturbative superpotentials from gaugino condensation. These are important in Section \ref{sec:KMS} once we discuss K\"ahler moduli stabilization. Finally, in Section \ref{sec:HDO} we discuss possibilities to describe the DBI action and Higgs-otic inflation in higher-derivative $\mathcal N = 1$ supergravity, extending the results of \cite{Bielleman:2016grv}.

%
\subsection{Higgs-otic inflation}
\label{sec:HI}

One of the original setups of \cite{Ibanez:2014swa} is a Type IIB compactification with $O3$/$O7$-planes and RR and NS three-form fluxes. These fluxes give rise to a monodromy potential for the position moduli of space-time filling D7-branes. In particular, the model features a compact orientifold with a local geometry of the form $(X \times T^2)/\mathbb Z_4$, where $X$ is some complex two-fold which is wrapped by a stack of D7-branes sitting at the singularity. In the example of \cite{Ibanez:2014swa}, the $U(N)$ gauge theory on the worldvolume of the branes contains the matter and Higgs sectors of the MSSM. Most importantly, some of the D7-branes may leave the singularity and travel through the bulk around the two-torus $T^2$ still satisfying tadpole cancellation conditions. When this happens a D-flat direction---lifted only by turning on fluxes---may give rise to inflation. The inflaton candidate in this case is the position modulus $\Phi$ of the mobile branes, which may be an MSSM Higgs boson.\footnote{If one drops the condition of identifying the inflaton with the Higgs boson, the case of a single moving D7-brane would be enough for the purposes of inflation. We will often use this simple case as a toy example to illustrate our results in the remainder of the paper, while commenting on the non-Abelian case as we go along.} The aforementioned three-form fluxes are the ISD components of $G_3$. We distinguish between the $(0,3)$-form flux $G=G_{\bar 1 \bar 2 \bar 3}$ and the $(2,1)$-form flux $S=\epsilon_{\bar 3 \bar j \bar k} G_{\bar 3 j k}$. The first class, $G$, breaks supersymmetry in the vacuum and gives rise to soft terms. The second class, on the other hand, gives rise to $\mu$-terms in the $\mathcal N=1$ superpotential. 

As we are interested in the dynamics of the fields on the branes, let us consider the corresponding Dirac-Born-Infeld (DBI) and Chern-Simons (CS) actions. In Einstein frame and the notation of \cite{Camara:2014tba} they read 
\begin{subequations}\label{eq:S1}
\begin{align}
S_\text{DBI} &= -\mu_7  \text{STr}\left( \int \text d^8x e^{-\phi}\sqrt{-\text{det}(P[e^{\phi/2}G_{MN}-B_{MN}])+\sigma F_{MN}}\right) \, , \\
S_\text{CS} &= \mu_7 \text{STr}\left( \int \text d^8x P[-C_6\wedge \mathcal{F}_2 + C_8]\right) \, ,
\end{align}
\end{subequations}
Here $M$ and $N$ are indices of the D7-brane worldvolume and $P$ denotes the pullback onto the same. The brane tension is given by $\mu_7=(2\pi)^{-7}(\alpha')^{-4}$, and $\sigma=2 \pi \alpha'$. The ten-dimensional space-time metric is denoted by $G_{MN}$, the NS two-form by $B_{MN}$, RR six- and eight-forms by $C_6$ and $C_8$, and the field strength of the gauge fields living on the brane by $F_{MN}$. We use the gauge-invariant combination $\mathcal{F}_2=B_{MN}-\sigma F_{MN}$ in the definition of the action only. For now we are not interested in worldvolume fluxes and matter fields coming from Wilson lines, so we ignore $F_{MN}$ in the following. Finally, STr denotes the symmetrized trace over gauge indices. The four-dimensional action for the position moduli can be obtained by performing the pullback of the metric, expanding the determinant, and integrating over the compact four-cycle wrapped by the brane. The brane worldvolume is parameterized by $\{x^{\mu}, x^m\}$ and the worldvolume fields only feel the local closed string background around the brane, so all quantities in \eqref{eq:S1} can be expanded in terms of the fluctuations of the transverse real fields $y^i=\sigma \varphi^i$ as follows,
\begin{align}
\text ds^2 &= Z(x^m)^{-1/2} \eta_{\mu\nu}\text d\hat{x}^\mu \text d\hat{x}^\nu + Z(x^m)^{1/2} \text ds^2_{\rm CY}\,,\nonumber\\
\tau &= \tau(x^m)= \tau_0 + \frac12 \sigma^2 \tau_{ij}\varphi^i\varphi^j + \dots \,, \label{back1}\\
G_3 &= \frac{1}{3!}G_{lmn}(x^p)\text dx^l\wedge \text dx^m\wedge \text dx^n\, , \quad G_{lmn}(x^p)= G_{lmn}+\dots\,,  \nonumber
\end{align}
to yield the desired action. Here $Z$ denotes the warp factor, $\tau = C_0 + i e^{-\phi}$ is the complex axio-dilaton, and $G_3 = F_3 - \tau H_3$ in terms of the usual RR and NSNS flux. Moreover, we define the complex scalar field that is the inflaton candidate as the following combination of the transverse coordinates, $\Phi=(\varphi^8+i\varphi^9)/\sqrt 2$. Instead of reproducing the results of \cite{Ibanez:2014swa}, we derive the four-dimensional action for $\Phi$ in Section \ref{sec:IASD} after including IASD flux.

%
\subsection{IASD flux and gaugino condensation}
\label{sec:IASD}

Since our goal is to stabilize all relevant closed string moduli in Higgs-otic inflation, we must consider extensions of the original setup in \cite{Ibanez:2014swa}, in which only ISD fluxes were included. Any further ingredient of the global compactification is parameterized in terms of the local background around the brane. Therefore, we must consider a local background rich enough to account for all further ingredients required to achieve moduli stabilization in our Type IIB compactification. One missing ingredient is non-perturbative superpotentials---required to stabilize, for example, K\"ahler moduli. In \cite{Baumann:2010sx} it was shown that such non-perturbative terms from gaugino condensation source IASD flux in the bulk of the compactification. Hence, in the following, we consider the additional contribution of $(1,2)$-form flux $D=\epsilon_{3 jk} G_{3 \bar j \bar k}$.\footnote{For a different kind of generalization that results in new sources of flattening for the D7-brane position potential, see \cite{Fernandotoappear}.}

The DBI action can then be written as follows. After performing the pullback, computing the trace, expanding the square root in powers of $\alpha' \partial_{\mu} \Phi$, and integrating over the wrapped four-cycle one obtains
\begin{align}\label{eq:DBI1}
\mathcal L_\text{DBI} = -  V_4 \mu_7 e^{\phi} f(B)(1+\sigma^2 Z \partial_{\mu}\Phi \partial^{\mu}\bar \Phi + ...) \,,
\end{align}
where $V_4$ is the volume of the wrapped four-cycle. We have assumed, for simplicity, that the internal four-cycle wrapped by the brane is Ricci flat so that the internal profile of the worldvolume fields is trivial. The ellipsis denotes higher-order terms in $\alpha' \partial_{\mu} \Phi$, which are sub-leading during slow-roll inflation. Moreover, 
\begin{align}
f(B)^2 
&=1 + \frac{1}{2Z} e^{-\phi} B_{ab}B^{ab} - \frac{1}{4Z^2} e^{-2 \phi} B_{ab}B^{bc}B_{cd}B^{da} + \frac{1}{8Z^2} e^{-2 \phi} (B_{ab}B^{ab})^2\,,\label{fB}
\end{align}
contains the scalar potential contribution from the DBI action to all orders in $\alpha'\Phi$, just like in \cite{Ibanez:2014swa}. Turning on the IASD flux $D$ does not introduce off-diagonal components in $B$, so we still find that \eqref{fB} completes to a perfect square, yielding
\begin{align}
f(B) &= 1 + \frac{1}{4 Z} e^{-\phi} B_{ab}B^{ab} \,.
\end{align}
The ten-dimensional Type IIB supergravity equations of motion relate the dilaton and the three-form fluxes of the global compactification. In particular, in the presence of both ISD and IASD fluxes, one obtains \cite{Camara:2014tba}
 \begin{align}
 \text{Im}(\tau_{3\bar 3}) &= -\frac{g_\text s}{4Z} (SD +  \bar S \bar D) \,,
 \end{align}
where we have again performed a local expansion of the dilaton field around the brane following~\eqref{back1},
\begin{align}
 e^{-\phi}  &= g_\text s^{-1}\left(1+ \sigma^2 \text{Im}(\tau_{3\bar 3})|\Phi |^2 + \frac12 \sigma^2 \text{Im}(\tau_{33}) \Phi^2 + \frac12 \sigma^2 \text{Im}(\tau_{\bar 3 \bar 3}) \bar \Phi^2 +\dots \right)\,.
\end{align}
Notice that $\tau_{33}$ and $\tau_{\bar 3 \bar 3}$ are not related to the fluxes, so they can be set to zero for simplicity. In a similar way we can use the equations of motion for the NS- and RR-forms,
\begin{align}
\text dB_2 &= -\frac{\textrm{Im}(G_3)}{\textrm{Im}(\tau)}\,, \\ \nonumber
\text dC_6&=H_3\wedge C_4 - *_{10}\, \textrm{Re}(G_3)\,, \label{bi}\\
\text dC_8&=H_3\wedge C_6-*_{10}\, \textrm{Re}(\tau)\,, \nonumber
\end{align}
 to write them in terms of $\Phi$ and the fluxes. The non-vanishing components are 
\begin{align}
B_{12} &= \frac{g_\text s \sigma}{2 i} \left[\bar G \Phi - (S- \bar D)\bar \Phi \right]\, , \\
(C)_{12} &= - \frac{g_\text s \sigma}{2 i Z} \left[\bar G \Phi - (S+ \bar D)\bar \Phi \right]\, , \\
(C)_{1\bar 1 2 \bar 2} &=  \frac{g_\text s^2 \sigma^2}{4 Z} \left( |G|^2 + |S|^2 - |D|^2] \right)|\Phi|^2 - \frac{g_\text s^2 \sigma^2}{4 Z} \left(GS \bar \Phi^2 + \bar G \bar S\Phi^2\right) \, .
\end{align}
Together with \eqref{eq:DBI1} this yields for the DBI action
\begin{align}\label{eq:DBI2}
\mathcal{L}_\text{DBI} &= - V_4 \mu_7 g_\text s (1+\sigma^2 Z \partial_\mu \Phi \partial^\mu \bar \Phi)\bigg[1+ \frac{g_\text s \sigma^2}{4Z} \Big(|G|^2+|S|^2+|D|^2)|\Phi|^2 \nonumber \\ 
&\ \ \  -  \bar G (\bar S - D )\Phi^2 - G ( S - \bar D )\bar \Phi^2 \Big) \bigg] \,.
\end{align}
Note that the negative constant contribution is cancelled by the orientifold contribution. From the CS action we find 
\begin{align}
\mathcal{L}_\text{CS} = \frac{V_4 \mu_7 g^2_\text s \sigma^2}{4Z}\left(-|\bar G \Phi - S \bar \Phi |^2 + |D|^2 |\Phi|^2  \right) \, .
\end{align}
Let us redefine $\Phi \to (V_4 \mu_7 g_\text s Z \sigma^2)^{-1/2} \Phi$ to obtain a canonical kinetic term at leading order in $\alpha'$. After combining both contributions we find the following kinetic terms and potential for $\Phi$,
\begin{subequations}\label{eq:DBICS1}
\begin{align}\label{eq:DBICS11}
\mathcal L_\text{kin} &= - \partial_\mu \Phi \partial^\mu \bar \Phi \left\{1 + \frac{1}{4 Z V_4 \mu_7} \left[ (|G|^2+|S|^2+|D|^2)|\Phi|^2 -\bar G (\bar S - D )\Phi^2 + \text{c.c.}\right] \right\}\\
V&=  \frac{g_\text s}{4 Z} \left(2 |G^*\Phi - S\bar \Phi |^2 + \bar G D \Phi^2 + G \bar D \bar \Phi^2\right)\, .
\label{eq:DBICS12}
\end{align}
\end{subequations}
which is exact to all orders in $\alpha'$ at two-derivative level. As in \cite{Ibanez:2014swa} the potential is quadratic in $\Phi$ and there is a nontrivial piece in the kinetic term which leads to a flattening of the effective inflation potential \cite{Bielleman:2016grv}. This piece is indeed proportional to the scalar potential from the DBI action. When $D$ vanishes, the scalar potential from the DBI action is equal to the CS contribution, so the correction to the kinetic term can be written as proportional to the full scalar potential itself. However, in the presence of a IASD flux $D$, both contributions are different and the correction to the kinetic term is only sensitive to the DBI part. In Section \ref{sec:DBICMB} we match this scalar potential with the supergravity calculation of the corresponding low-energy $\mathcal N =1$ theory after K\"ahler moduli stabilization. We then use the result for the kinetic term derived from the DBI action to analyze the flattening of the effective potential when combined with K\"ahler moduli stabilization. 

%
\subsection{The DBI action and higher-derivative supergravity}
\label{sec:HDO}

In order to describe Higgs-otic inflation and moduli stabilization in a unified framework it is desirable to have an $\mathcal{N} = 1$ supergravity description of the DBI + CS action in \eqref{eq:DBICS1}. In the case with only ISD fluxes, $D=0$, this can be done along the lines of \cite{Bielleman:2016grv}. We have shown above that, in this case, the brane worldvolume action yields a Lagrangian for $\Phi$ of the schematic form
\begin{align}
\mathcal{L}=-[1+a V(\Phi)] \partial_\mu \Phi \partial^\mu \bar \Phi - V(\Phi) + \mathcal{O}[(\partial_\mu \Phi)^4] \, ,
\label{eq:effactiond7}
\end{align}
where $a\propto 1/(V_4 \mu_7 g_\text s)$. As in Section \ref{sec:IASD}, the scalar potential is exact in $\alpha'$, since $B$ is diagonal, and \eqref{eq:effactiond7} only receives higher-order derivative corrections that are negligible in the slow-roll regime. In \cite{Bielleman:2016grv} we have argued that \eqref{eq:effactiond7} can be written in terms of standard $\mathcal N =1$ data if one allows the K\"ahler potential to depend on superspace derivatives of the chiral superfields such that it includes the following term
\begin{align}
\delta K \sim \frac{1}{\Lambda^4} |\Phi + \bar \Phi|^2 \partial_\mu \Phi \partial^\mu \bar \Phi \, .
\end{align}
Here $\Phi$ denotes the superfield containing the complex scalar and  $\Lambda$ is the cut-off scale of the supergravity effective field theory. In the string compactifications discussed here, this is proportional to the string scale $M_\text s = (\alpha')^{-1/2}$. After performing the superspace integral in the global limit, one finds a correction to the kinetic term of the scalar $\Phi$ which is proportional to the potential, as advertised in \eqref{eq:effactiond7}:\begin{equation}
\delta \mathcal{L} \sim -\frac{8 }{\Lambda^4}  |F|^2 \partial_\mu \Phi \partial^\mu \bar{\Phi}\,,
\label{eq:global}
\end{equation}
and no corrections to the scalar potential arise. In addition to this desired term, the theory also features terms with derivatives of the auxiliary field $F$. These can be consistently ignored below the cut-off $\Lambda$, for more details see \cite{Baumann:2011nm,Bielleman:2016grv}. 

Since inflation is, by definition, a period of quasi de Sitter expansion of the early universe, consistency of this approach requires an embedding of the previous idea into supergravity, i.e., supersymmetry in curved backgrounds. The general theory of coupling a global K\"ahler- and superpotential to gravity is well understood, see \cite{Wess:1992cp} for a review. Specifically, the curved-space supergravity Lagrangian involving $K$ and $W$ reads
\begin{align}
\mathcal{L} &= \int \text d^2 \Theta \ \mathcal{E}\left[ \frac38(\mathcal{\bar D}-8R) e^{K(\Phi, \bar \Phi)} + W(\Phi) \right] + \text{h.c.} \, ,
\end{align}
in the conventions of \cite{Wess:1992cp}. Here $\mathcal{E}$ is the chiral density and $(\mathcal{\bar D}-8R)$ is the chiral projector that ensures that the integral over superspace gives a supersymmetric Lagrangian. In this expression, $R$ is the supergravity superfield containing the supergravity multiplet $(\mathcal{R}, b_\mu, M)$. Just like in  flat space, we can allow for $K$ to depend on derivatives of the chiral superfield $\Phi$ and in that way generate the non-standard kinetic terms we are interested in. This technique was extensively applied in a systematic study of higher-derivative operators in supergravity in \cite{Ciupke:2016agp}, whose results we use to find the component Lagrangian following from the K\"ahler potential\footnote{Compared to the K\"ahler potential given in \cite{Bielleman:2016grv} we have absorbed a factor of $(S+\bar S)(U+ \bar U)$ in the definition of $\Lambda$ for convenience.}
\begin{align}
K &= -\log\left[(S+\bar S) (U + \bar U) - \frac12 (\Phi + \bar \Phi)^2\right] -3 \log[T + \bar T]\,+ \delta K, 
\label{eq:Kpotential}
\end{align}
\begin{equation}
\delta K=\frac{\mathcal{T}}{\Lambda^4}\partial_\mu  \Phi \partial^\mu \bar \Phi\,, \qquad \mathcal{T}=\frac{|\Phi+\bar{\Phi}|^2}{6}.
\label{eq:deltaK}
\end{equation}
As discussed in more detail in \cite{Bielleman:2016grv}, \eqref{eq:Kpotential} and \eqref{eq:deltaK}, when supplemented by the superpotential 
\begin{align}\label{eq:supo17}
W = W_0 + \mu \Phi^2\,,
\end{align}
lead to the supergravity embedding of \eqref{eq:effactiond7}. In accordance with the notation used in \cite{Ibanez:2014swa,Bielleman:2016grv} $S$ denotes the axio-dilaton, $U$ is a complex structure modulus, $T$ is a volume modulus, and $\Phi$ denotes the D7-brane position modulus of Higgs-otic inflation, as discussed in Section \ref{sec:HI}. Let us postpone the discussion of moduli stabilization until Sections 3 and 4, and only focus on the multiplet $\Phi$ and its kinetic term. Meanwhile, we treat $S$, $T$, and $U$ as constant.

Note that, unlike the globally supersymmetric case with only the single mass scale $\Lambda$, once we couple to gravity we are bound to deal with two cut-off scales: $\Lambda$ and $M_\text p$. In the following, it is crucial to keep track of both of them as they determine the relevant terms in $\mathcal{L}$. With this in mind we reinstate the factors of $M_\text p$ in the results of \cite{Ciupke:2016agp} by noting that the mass dimension of $M$ and $b_\mu$ is three and that of the curvature, $\mathcal D_\mu$, and $F$ is two.  Following \cite{Ciupke:2016agp} we write the component expansion of \eqref{eq:deltaK} in the Jordan frame as follows,\footnote{The standard derivation of the two-derivative supergravity Lagrangian leads to a gravitational coupling of the form $e^{K/3} \mathcal{R}$, cf.~\cite{Wess:1992cp}. This is the frame we choose for the moment.}
\begin{align}
\delta \hat{\mathcal{L}}/ \sqrt{|g|} = - \frac{1}{2} \Omega \mathcal{R}  - \delta V_{} + \mathcal{L}^{(\text{4-der})} + \mathcal{L}^{(\text{2-der})}    \ ,
\end{align}
where
\begin{align}  
\Omega &=  \frac{4 \mathcal{T}}{\Lambda^4} \lvert \partial \Phi \rvert^2 \label{eq:deltaK1} \\\
\delta V &= - \frac{4 \mathcal T}{3 \Lambda^4 M_\text p^4}  \lvert F \rvert^2 \lvert M \rvert^2 \, , \\
 \Lambda^4 \mathcal{L}^{(\text{2-der})}  &=  \mathcal{T}_{\bar{\Phi}} \left[ \frac{1}{M_\text p^2}MF (\partial_\mu \bar{\Phi})^2 +\frac{1}{M_\text p^2} \bar{M} \bar{F} \lvert  \partial_\mu \Phi \rvert^2 - 6 \bar{F} \partial_\mu F \partial^\mu \bar{\Phi} - \frac{1}{M_\text p^2} 4i \lvert F \rvert^2 b^\mu \partial_\mu \bar{\Phi} \right] \label{eq:deltaK2}\\
   & \quad - 3 \mathcal{T}_{\Phi\bar{\Phi}} \lvert F \lvert^2 \lvert  \partial_\mu \Phi \rvert^2  - \mathcal{T} \left( \frac{1}{3 M_\text p^4} \lvert  \partial_\mu \Phi \rvert^2 \lvert M \rvert^2 + \frac{4}{3 M_\text p^4} \lvert F \rvert^2 b_a b^a + 3 \lvert \partial_\mu F \rvert^2 \right) \nonumber \\
   & \quad  + \mathcal{T}\left(\frac{1}{M_\text p^2} FM \Box \bar{\Phi} +\frac{1}{M_\text p^2} M \partial_\mu F \partial^\mu \bar{\Phi} - \frac{1}{M_\text p^2}F \partial_\mu M \partial^\mu \bar{\Phi}\right) \nonumber \\
   & \quad + \frac{4}{3} \mathcal{T} i b^\mu\left(\frac{1}{M_\text p^4} FM \partial_\mu \bar{\Phi} +\frac{3}{M_\text p^2} \bar{F} \partial_\mu F\right) + \text{h.c.} \nonumber \, ,
\end{align}
and
\begin{align}
 \Lambda^4 \mathcal{L}^{(\text{4-der})} &= -3 \mathcal{T}_{\bar{\Phi}} \left[ \lvert  \partial_\mu \Phi \rvert^2 \left(\Box \bar{\Phi} + \frac{2}{3 M_\text p^2} i b^\mu \partial_\mu \bar{\Phi} \right) +\frac{2}{M_\text p^2} \partial^\mu \bar{\Phi} \, \partial^\nu \Phi \, \mathcal{D}_\mu \mathcal{D}_\nu \bar{\Phi} \right]  \label{eq:deltaK3}\\
   & \quad -3 \mathcal{T}_{\Phi\bar{\Phi}}\lvert  \partial_\mu \Phi \rvert^2 (\partial_\mu \bar{\Phi})^2 + 3  \mathcal{T}   \partial^\mu \Phi \partial^\nu \bar{\Phi} \left[ \mathcal{R}_{\mu\nu} + \frac{2}{9 M_\text p^4} b_\mu b_\nu + \frac{2}{3 M_\text p^3} i \mathcal{D}_\nu b_\mu \right] \nonumber \\
   & \quad   - 3  \mathcal{T} \partial_\mu \Phi \, \mathcal{D}^\mu \left(\frac{1}{M_\text p} \Box \bar{\Phi} +\frac{2}{3 M_\text p^3}i b^\nu \partial_\nu \bar{\Phi}\right)  + \text{h.c.}\nonumber \, .
\end{align}
Two facts stand out in this contribution to the action. First, there is a correction to the scalar potential proportional to $|M|^2|F|^2$. This is noteworthy because one of the guiding principles in the determination of the K\"ahler potential in \cite{Bielleman:2016grv} was the absence of corrections to $V$. There is no contradiction with our previous result, though, since $\delta V \to 0$ in the rigid limit $M_\text p \to \infty$, where the result matches the one from the DBI action. Second, there are nonminimal couplings between the scalar $\Phi$ and the Ricci tensor and scalar. While the coupling to the Ricci scalar can be dealt with via a simple Weyl rescaling, that is not the case for the coupling to $\mathcal{R}_{\mu\nu}$. 

Let us count dimensions to determine which of these terms dominate in the action. In the component expansion in \eqref{eq:deltaK1}--\eqref{eq:deltaK3} we find operators up to dimension  twelve, with the following suppressions: $1/\Lambda^4$, $1/(\Lambda^4 M_\text p^2)$, and $1/(\Lambda^4 M_\text p^4)$. Since in the standard $\mathcal{N}=1$ supergravity action one has terms of the order $1/M_\text p^n$, with $n=0,2,4$, we focus on the terms up to mass dimension eight. This truncation is justified since, at higher-order, terms in $\mathcal L$ can be sourced by higher-order corrections to $K$ which we do not consider. Moreover, as in the flat-space case, let us consider $\partial_\mu F=0$, since the dynamics of $F$ are an artifact of the effective field theory description. Then only three terms survive in the higher-derivative correction of the Lagrangian,
\begin{align}
\Lambda^4 \delta \hat{\mathcal{L}} / \sqrt{|g|}= -\frac{1}{3}\mathcal{R}(\Phi + \bar \Phi)^2 |\partial_\mu \Phi|^2 - |F|^2 |\partial_\mu \Phi|^2 + \frac12 (\Phi+\bar \Phi)^2 \partial^\mu \Phi\partial^\nu \bar{\Phi}\mathcal{R}_{\mu\nu} \,.
\label{eq:LHDsimp}
\end{align}
Notice that we have absorbed overall coefficients of $\mathcal O(1)$ and a possible constant $a$ into the definition of $\Lambda$. As a result, the inclusion of the correction $\delta K$ yields the same result as in global supersymmetry, plus nontrivial curvature couplings. And even these two additional terms are irrelevant in the model we consider: both are proportional to $\mathcal T$ and thus to $\text{Re}(\Phi)$, which means they vanish in the model of Higgs-otic inflation considered in this paper. We will see that, in order to ensure stability once K\"ahler moduli stabilization is taken into account, only the lightest field $\text{Im}(\Phi)$ is excited during inflation, while $\text{Re}(\Phi)$ remains stabilized at the origin. We note however that the effect of such nonminimal couplings on the quantum fluctuations of the system during inflation is a subtler issue that requires further study.

We can now recast the action into its most useful form via a conformal transformation to Einstein frame, leading to 
\begin{align}
\mathcal{L}/\sqrt{|g|}=\mathcal{L}_0 - \frac{1}{\Lambda^4}  e^{K} |F|^2|\partial_\mu \Phi|^2 \ ,
\label{eq:LFinal}
\end{align}
where 
\begin{align}
 \mathcal{L}_{(0)} / \sqrt{|g|} = &- \frac12 M_\text p^2 \mathcal{R} - K_{\Phi\bar{\Phi}} \, \partial_{\mu} \Phi \partial^{\mu} \bar{\Phi}  -V_{(0)} \ ,
 \label{eq: L0}
\end{align}
is the usual supergravity Lagrangian, with $V_0$ being the F-term potential. Note again that \eqref{eq:LFinal} is much simpler than our starting point, and is essentially the global result of \eqref{eq:global}. If, as mentioned above, we choose a superpotential that leads to a quadratic scalar potential as in the DBI and CS actions, we can find the on-shell kinetic terms using \eqref{eq:supo17},
\begin{align}
\mathcal{L}_\text{kin} &= - \left (K_{\Phi \bar \Phi} + \frac{1}{\Lambda^4} e^{K} |F|^2 \right)|\partial_\mu \Phi|^2   \\
&= - \left(\frac12 +3a  \frac{s \mu^2 \varphi^2}{8 t_0^3}\right)(\partial_\mu \varphi)^2,
\label{eq:kinfinnaive}
\end{align}
where $\varphi = \sqrt {2/s} \ \text{Im}(\Phi)$, $s = \langle(S+\bar S)(U+ \bar U)\rangle$, and $t_0 = \langle T \rangle$. 
To summarize, we have shown that the DBI and CS actions with ISD flux can be effectively described by the K\"ahler potential in \eqref{eq:Kpotential} and \eqref{eq:deltaK}. Coupling to gravity does not, in the end, make the Lagrangian more complicated as long as the cut-off scale is much larger than the dynamical scale of inflation, and $\text{Re}(\Phi) = 0$ during inflation. 

A crucial assumption in the above derivation is the absence of IASD fluxes, which induce additional terms in the DBI and CS actions. As we will see shortly, the inclusion and stabilization of moduli fields, in particular K\"ahler moduli, calls for the inclusion of non-perturbative effects which act as IASD flux in the bulk. Consequently, the supergravity embedding of this more complex system requires a deviation from the ideas presented in this section, making the identification of the interesting operator(s) a significantly more arduous task. In Section \ref{sec:DBICMB1} we discuss this is more detail and present an alternative method to tackle this problem.

%

\section{K\"ahler moduli stabilization and backreaction}
\label{sec:KMS}

In this section we extend the supergravity setup of Higgs-otic inflation by an explicit treatment of K\"ahler moduli stabilization and its backreaction on the inflationary physics. As an instructive toy model, we focus on the stabilization of a single volume mode $T$ via the setup of KKLT \cite{Kachru:2003aw}. After briefly reviewing the KKLT mechanism and the parameters involved, we compute the backreaction of the stabilized K\"ahler modulus on the inflationary scalar potential of Higgs-otic inflation, following the analysis in \cite{Buchmuller:2015oma}. Then we combine these results with those of Section \ref{sec:DBI} to account for the nontrivial kinetic term from the DBI action. 

%
\subsection{The KKLT mechanism -- a brief recap}
\label{sec:KKLT}

Let us consider the stabilization of a single K\"ahler modulus $T$. In the original setup of \cite{Kachru:2003aw}
one has 
\begin{align}\label{eq:KKLT}
K = -3 \log\left[T + \bar T\right]\,, \quad W = W_0 + A e^{-\alpha T}\,,
\end{align}
where $W_0$ is the vacuum expectation value of the Gukov-Vafa-Witten flux superpotential, and the non-perturbative term is sourced either by a Euclidean D3-brane instanton or a gaugino condensate on a stack of D7-branes. The scalar potential
\begin{align}
V = e^K \left( K^{T \bar T} D_T W \overline{D_T W} - 3|W|^2 \right)\,,
\end{align}
has two extrema, one of which is an AdS minimum defined by $D_T W = 0$. To uplift this vacuum to one with vanishing cosmological constant, one may add the term
\begin{align}
V_\text{up} = e^K \Delta^2
\end{align}
to the potential.\footnote{The microscopic source of the uplift is unimportant in our case. It could be, for example, F-terms of matter fields or an anti-D3-brane in a strongly warped region of the compactification.} One then finds a Minkowski minimum by solving $\partial_T V = V = 0$. The modulus is stabilized at $\langle \text{Im}(T)\rangle = 0$ and $\langle \text{Re}(T) \rangle \equiv t = t_0$. In the vacuum we find the following relations among the parameters,
\begin{align}\label{eq:solvac}
A = -\frac{3W_0  (\alpha t_0-1)e^{\alpha t_0}}{2 \alpha t_0 (\alpha t_0+2)-3}\,, \qquad \Delta^2 = \frac{12 \alpha^2 t_0^2 W_0^2 (\alpha t_0-1) (\alpha t_0+2)}{[3-2 \alpha t_0 (\alpha t_0+2)]^2}\,.
\end{align}
The first equality defines $t_0$ in terms of the parameters in $W$. In this vacuum, the auxiliary field of $T$ breaks supersymmetry, and
\begin{align}
F_T \equiv e^{K/2} \sqrt{K^{T \bar T}} D_T W = \frac{3 \sqrt 3 W_0}{4 \sqrt 2 \alpha t_0^{5/2}} + \mathcal{O} \left[(\alpha t_0)^{-2} \right]\,. 
\end{align}
Notice that we have expanded in inverse powers of $\alpha t_0$. This is a good expansion parameter because $\alpha t_0 \gg 1$ for consistency of the single-instanton approximation made in \eqref{eq:KKLT}. Analogously we find 
\begin{align}
m_{3/2} = e^{K/2} W = \frac{W_0}{(2 t_0)^{3/2}} + \mathcal{O} \left[(\alpha t_0)^{-1} \right]\,,
\end{align}
in the vacuum. The mass of the canonically normalized modulus is 
\begin{align}
m_{t} = 2 \alpha t_0 m_{3/2} \,.
\end{align}

A few comments are in order. First, the Minkowski vacuum is metastable. It is separated from the runaway minimum at $T = \infty$ by a barrier whose height is approximately the same as the depth as the original AdS vacuum, i.e., $V_\text{barrier} \simeq 3 m_{3/2}^2$. This is of great importance later on, when the Lagrangian of the modulus is coupled to inflation. Second, the scale of the parameters in $W$ is somewhat constrained: since $t_0 \gg 1$ is required by the supergravity approximation, according to \eqref{eq:solvac} it must be $W_0 \ll 1$ as long as $A \sim \mathcal O(1)$.

%
\subsection{Backreaction and effective potential}
\label{sec:back}

We are now in a position to discuss moduli stabilization in the supergravity model of Higgs-otic inflation. Before we turn to the more complicated non-Abelian setup involving mobile D7-branes, let us consider the simple but instructive case of a single brane. 

\subsubsection{A single-field toy model}
\label{sec:single}

In this case there is a single complex scalar field, $\Phi$, parameterizing the motion of the brane away from the orientifold singularity, as described in Section \ref{sec:HI}. Following \cite{Ibanez:2014swa,Bielleman:2016grv} the corresponding $\mathcal N = 1$ supergravity theory---postponing further discussion of higher-derivative corrections until Section \ref{sec:DBICMB}---is then defined by 
\begin{subequations}\label{eq:toy1}
\begin{align}\label{eq:toy1K}
K &= - \log \left[(S + \bar S)(U + \bar U) - \frac12 (\Phi + \bar \Phi)^2  \right] - 3 \log \left[T + \bar T\right]\,, \\
W &= \mu \Phi^2 + W_0 + A e^{- \alpha T}\,,\label{eq:toy1W}
\end{align}
\end{subequations}
where $S$ and $U$ denote the axio-dilaton and a complex structure modulus, respectively.\footnote{Notice that we assume that a possible dependence of the non-perturbative piece on the open-string modulus is negligible. Cf.~\cite{Ruehle:2017one} for a discussion of this issue in similar large-field models.} Such a setup arises, for example, when the complex two-fold $X$ is a four-torus, two of the three complex structure moduli are assumed to be stabilized, and the three K\"ahler moduli of the tori are identified. Moreover, in this section we are only interested in the backreaction of the K\"ahler modulus $T$, so we assume $S$ and $U$ to be stabilized supersymmetrically at a high scale. We come back to this issue in more detail in Section \ref{sec:CS}. Denoting the product of the vacuum expectation values of $U$ and the dilaton by $s$, we hence work with the effective K\"ahler potential 
\begin{align}\label{eq:toy1Keff}
K = - \log \left[s - \frac12 (\Phi + \bar \Phi)^2  \right] - 3 \log \left[T + \bar T\right]\,.
\end{align}
The model defined by \eqref{eq:toy1W} and \eqref{eq:toy1Keff} is nearly identical to the one considered in \cite{Buchmuller:2015oma}. As a warm-up for the more complicated non-Abelian case, let us outline how the stabilization of $T$ works during inflation, and how the modulus back-reacts on the effective potential of the inflaton.

\paragraph{Interaction during inflation} 
$\,$\\
During inflation there are terms coupling $T$ and $\Phi$ in the Lagrangian. More specifically, for real superpotential parameters only $\text{Re}(T)$ and $\text{Im}(\Phi)$ interact; the volume modulus and the inflaton field.\footnote{The more general case of complex $\mu$ and $W_0$ was considered in \cite{Buchmuller:2015oma}. By field redefinitions and K\"ahler transformations one can always rotate to a frame where the results are qualitatively the same as in the case considered here.} Both the axion of $T$ and $\text{Re}(\Phi)$ are stabilized at the origin with a large mass, so we can safely neglect them in the following, cf.~\cite{Buchmuller:2015oma} for details. The full scalar potential reads 
\begin{align}
V(t, \varphi) = \frac{1}{8 s t^3} \left[ \Delta^2 + \frac43 \alpha t A e^{-2 \alpha t} \left(3 A + \alpha t A + 3 W_0 e^{\alpha t}\right) + 2 \left(-\alpha t A e^{-\alpha t}+ s \mu \right) s \mu \varphi^2 \right]\,,
\end{align}
where $\varphi$ is the canonically normalized inflaton field and $t$ denotes the real part of $T$. The interaction terms between $t$ and $\varphi$ imply that, even if $t$ is much heavier than $\varphi$, during inflation the minimum of the modulus potential is inflaton-dependent. We assume that $t$ traces its minimum adiabatically---an assumption which is satisfied as long as a large mass hierarchy is present. Integrating out $t$ at its $\varphi$-dependent value then leads to additional terms in the effective potential for $\varphi$. This is what we refer to as ``backreaction" of the modulus field.

After imposing \eqref{eq:solvac} to eliminate $A$ and $\Delta$ it is useful to expand the potential in terms of $t = t_0 + \delta t(\varphi)$, where $t_0$ denotes the modulus expectation value after inflation. We are thus treating inflation as a perturbation of moduli stabilization. Unfortunately, we cannot minimize $V$ and solve for $\delta t(\varphi)$ analytically to all orders. So we expand $V(t,\varphi)$ up to second order in $\delta t(\varphi)$ and minimize afterwards. The result reads
\begin{align}\label{eq:deltat}
\frac{\delta t (\varphi)}{t_0} = \frac{s \mu \varphi^2}{2 \alpha t_0 W_0} + \mathcal O (H^2/m_t^2)\,.
\end{align}
As stressed in \cite{Buchmuller:2015oma} one must have $m_t > H$ throughout the inflationary period to guarantee stability of $T$. Therefore, it is instructive to expand all relevant quantities in powers of $\mu/W_0$, as we have done above. For consistency of the expansion around $t_0$ we must demand that \eqref{eq:deltat} is small compared to one. This leads to an important constraint on the parameters of the superpotential,
\begin{align}\label{eq:constraint}
\frac{W_0}{\mu} > \frac{s \varphi^2}{2 \alpha t_0}\,.
\end{align}
As one can verify explicitly, for example, by solving the full equations of motion numerically or by simply drawing the full potential, $\delta t (\varphi) = 1$ is the point where the modulus is lifted over its KKLT barrier and the theory decompactifies. Hence, the theory is only well-behaved as long as $\delta t (\varphi) \ll t_0$.

\paragraph{Effective inflaton potential} 
$\,$\\
Now, while this is the case and \eqref{eq:constraint} is satisfied we can integrate out the modulus by inserting \eqref{eq:deltat} into $V(t,\varphi)$. This yields the Wilsonian effective action for the inflaton field. The resulting scalar potential reads, to leading order in $\alpha t_0$ and $H/m_t$,
\begin{align}\label{eq:veff}
V_\text{eff}(\varphi) = \frac{1}{4 t_0^3} \left( s \mu^2 \varphi^2 + \frac32 \mu W_0 \varphi^2 - \frac38 s \mu^2 \varphi^4 \right)  + \dots \,.
\end{align}
Let us consider this expression for a moment. The first term is the supersymmetric mass term for $\varphi$ which we naively expected, and which is the term driving inflation in \cite{Ibanez:2014swa}. The third term is surprising at first sight: its stems from the $-3 |W|^2$ piece of the supergravity scalar potential, which we thought we had gotten rid of by including the no-scale field $T$ in the theory. This is what the backreaction of $T$ introduces. The interaction terms between $t$ and $\varphi$ in the Lagrangian interfere with the no-scale cancellation during inflation. This tells us that ``neglecting'' moduli stabilization in string theory models of large-field inflation is a very dangerous thing. What potentially saves the theory at hand is the second term in \eqref{eq:veff}. In the regime required by \eqref{eq:constraint}, i.e. $W_0 \gg \mu$, it is bigger than the first term and can drive inflation. 

Note that the second and third term in \eqref{eq:veff} only arise after minimizing with respect to $T$. They vanish in case the non-perturbative term in \eqref{eq:toy1W} is absent. This can be seen very clearly for the new mass term in case we use the equations of motion for $T$ to eliminate the parameter $W_0$ instead of $A$ in \eqref{eq:solvac}. We then obtain
\begin{align}
V_\text{eff}(\varphi) = \frac{1}{4 t_0^3} \left( s\mu^2 \varphi^2 - \alpha t_0 A e^{-\alpha t_0} \mu \varphi^2 + \dots \right) + \dots\,.
\end{align}
Clearly, the new dominant mass term vanishes if $A = 0$, in which case $V(t)$ has no minimum.

Neglecting the supersymmetric mass term proportional to $\mu^2$, we can write the relevant potential as follows, 
\begin{align}\label{eq:veff2}
V_\text{eff}(\varphi) = \frac{3}{8 t_0^3} \mu W_0 \varphi^2 \left(1 - \frac{s \mu}{4 W_0} \varphi^2 \right)  + \dots \,.
\end{align}
In a sense, this is a quadratic potential with a correction term scaling as $H/m_{3/2}$ or $H/m_{t}$, as naively expected. $V_\text{eff}$ has a maximum at $\varphi_\text{c}^2 = 2 W_0/s \mu$. Because we must require that $\varphi_\star < \varphi_\text{c}$ for inflation to be successful, this leads to a parameter contraint
\begin{align}\label{eq:constraint2}
\frac{W_0}{\mu} > \frac{s \varphi_\star^2}{2}\,,
\end{align}
which is slightly tighter than the one in \eqref{eq:constraint}. As mentioned above, this corresponds to a flux tuning of the superpotential parameters. We discuss the viability of this tuning in Section 5.

\begin{figure}[t]
\centering
\includegraphics[width=0.6\textwidth]{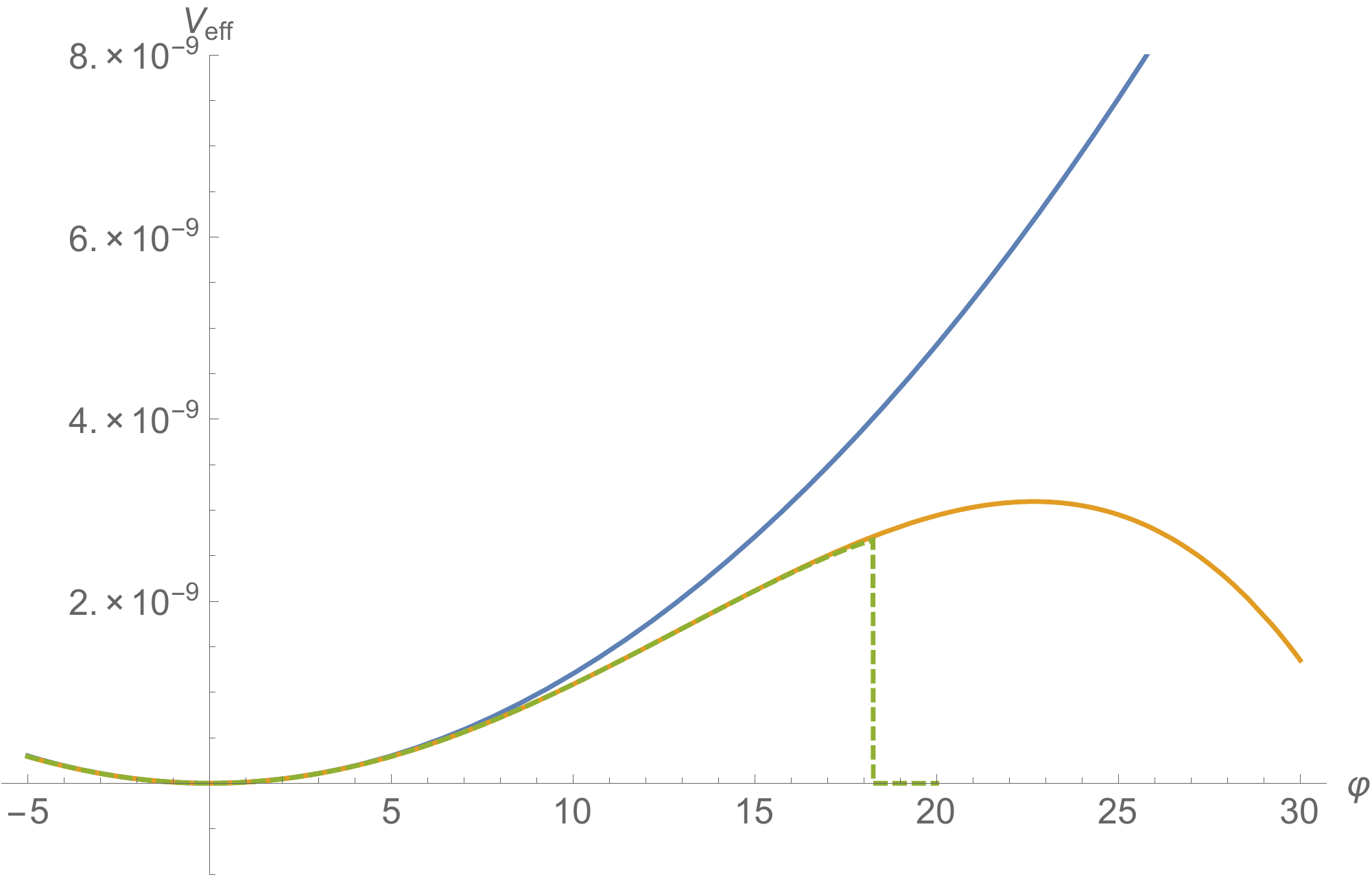} 
\caption{Effective inflaton potential obtained analytically via the second-order expansion in $\delta t(\varphi)$ (orange line) and numerically to all orders (green dashed line), in comparison with the naive quadratic potential (blue line). The flattening effect of integrating out $T$ is evident. The orange curve is obtained from the result \eqref{eq:veff2} with all higher-order terms in $(\alpha t_0)^{-1}$ taken into account.
\label{fig:veff2}}
\end{figure}

\paragraph{A benchmark model} 
$\,$\\
Let us finally consider a parameter example. Remember that it should satisfy both \eqref{eq:constraint2} and correct normalization of the scalar perturbations on the would-be inflationary trajectory. In Figure \ref{fig:veff2} we have displayed the effective inflaton potential for the following parameter choice,
\begin{align}\label{eq:toypar1}
W_0 = 0.005\,, \quad  \mu = W_0/400\,, \quad s = 1 \,, \quad t_0 = 10\,, \quad \alpha = 2 \pi/5\,,
\end{align}
in comparison with a purely quadratic potential as obtained in \cite{Ibanez:2014swa} (blue line). The leading-order effective potential (orange line) has a local maximum at $\varphi_\text{c} \approx 23$. It is no surprise that the position of the maximum is very close to the point where $T$ is destabilized and $\delta t (\varphi) \approx 1$. This point, once more, signalizes the breakdown of the effective theory we have obtained after integrating out $T$. Beyond $\varphi_\text{c}$ the modulus can no longer be integrated out and we obtain a very undesirable theory which decompactifies. This is clear after considering the green dashed line in Figure \ref{fig:veff2}: it is the effective inflaton potential after integrating out $T$ to all orders numerically.\footnote{Notice the good agreement with the analytic result obtained from the second-order expansion of the potential.} The point where the curve drops is the point where the minimum in the modulus direction disappears and the theory decompactifies. However, to the left of the maximum value 60 $e$-folds of inflation may take place, as discussed in \cite{Buchmuller:2015oma}. We return to the details of the inflationary phase in Section \ref{sec:DBICMB}. 

\begin{figure}[t]
\centering
\includegraphics[width=0.6\textwidth]{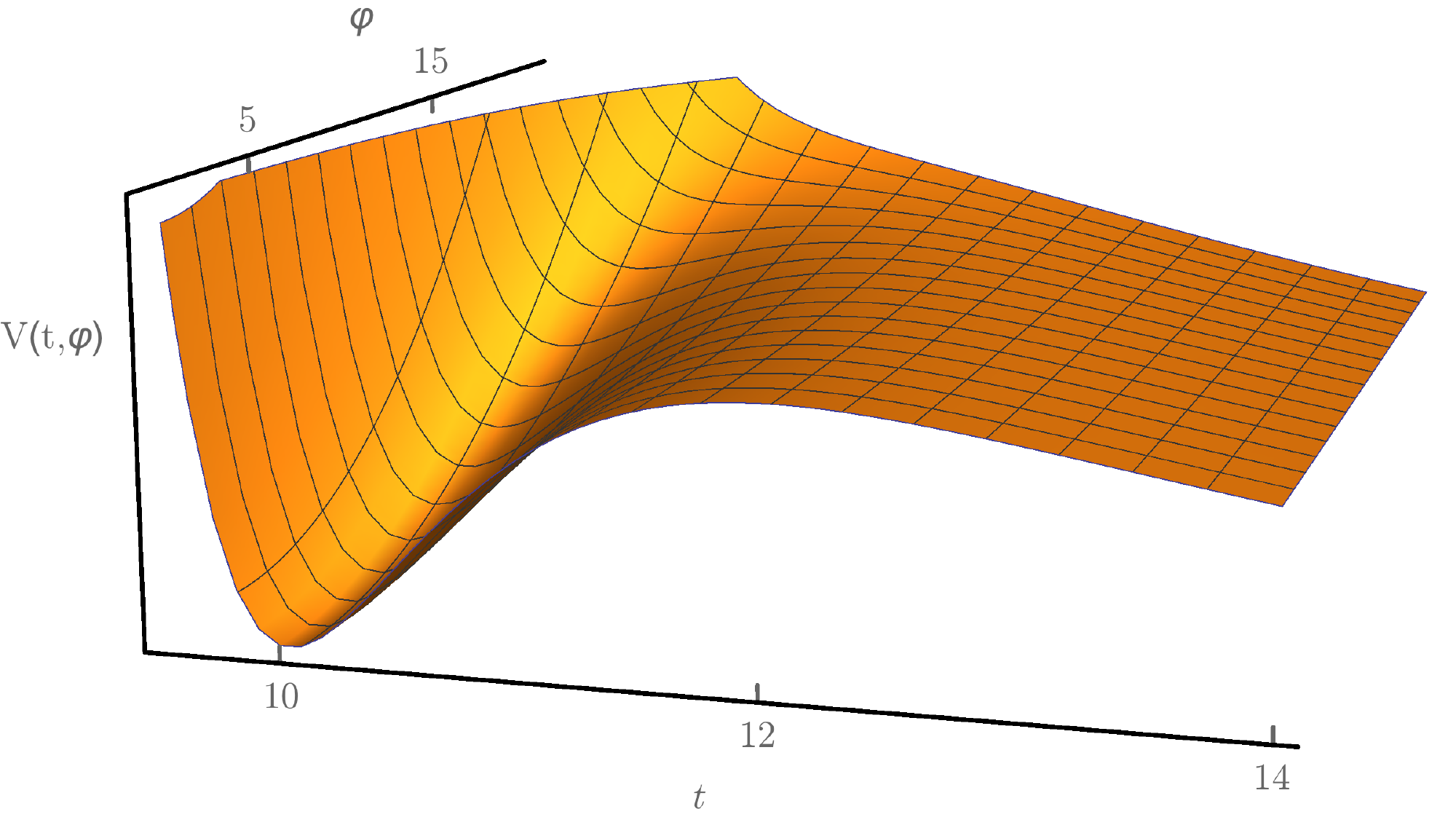} 
\caption{Scalar potential in the $t-\varphi$ plane. Evidently, the initial conditions must be very fine-tuned to allow for 60 $e$-folds of slow-roll inflation without destabilizing $t$.
\label{fig:4}}
\end{figure}

Last but not least, let us consider the full scalar potential in the $t -\varphi$ plane. This is shown in Figure \ref{fig:4} for the same parameter values as above. It clearly shows the flat valley along the minimum of $T$ in which slow-roll inflation can take place. However, it also highlights the amount of fine-tuning of initial conditions that is necessary to allow for inflation without destabilization of $T$. Of course, the necessary amount of fine-tuning can be reduced by increasing the tuning between $W_0$ and $\mu$, which pushes $\varphi_\text c$ to larger values.

\subsubsection{A realistic setup with MSSM Higgs multiplets}
\label{sec:2Higgs}

As pointed out in \cite{Ibanez:2014swa,Bielleman:2015lka} the inflationary scalar potential in the original Higgs-otic inflation model is more complicated than in the toy model of Section~\ref{sec:single}. In Higgs-otic inflation, a stack of six D7-branes is placed at an orbifold $Z_4$ singularity, so that the orbifold action gives rise to the gauge group and spectra of the SM. The position modulus $\Phi$ of the stack of branes transforms in the adjoint representation of the non-Abelian gauge group. As explained in \cite{Ibanez:2014swa}, one of the D7-branes and its orbifold image can leave the singularity and travel through the transverse torus. The off-diagonal fluctuations of the transverse field $\Phi$ correspond then to the MSSM Higgs doublets $H_\text u$ and $H_\text d$ and can yield inflation. Following \cite{Ibanez:2014swa}, we should consider the string-effective action defined by
\begin{subequations}\label{eq:toy3}
\begin{align}
K &= - \log \left[s - \frac12 (H_\text u + \bar H_\text d) (\bar H_\text u + H_\text d)  \right] - 3 \log \left[T + \bar T\right]\,, \\
W &= \mu H_\text u H_\text d + W_0 + A e^{-\alpha T}\,.
\end{align}
\end{subequations}
In principle, we thus have to deal with a complex modulus field and two real scalars from the inflationary sector parameterizing the two flat directions on the transverse torus. This makes the analysis a bit more involved. Technically it works the same way as above, though. The strategy is to expand around $t = t_0 + \delta t (H_\text u, H_\text d)$, then minimize, and insert the result for $\delta t$ to obtain the effective inflaton potential. 

\paragraph{Original potential without moduli stabilization} 
$\,$\\
The approach of \cite{Ibanez:2014swa} was to neglect the $T$-dependent piece in $W$, to neglect the dynamics of $T$ entirely, and to assume that $D_T W \neq 0$ which leads to a perfect no-scale cancelation of the term proportional to $-3|W|^2$. This yields the following positive definite scalar potential,\footnote{Note the appearance of a few factors of 2 compared to the potential in \cite{Ibanez:2014swa}. These are due to relative factor in the K\"ahler potential used here.}
\begin{align}\label{eq:v4}
V = \frac{1}{8 s t_0^3} \left[ (W_0^2 + 2 s \mu W_0 + 2 s^2 \mu^2) (|H_\text u|^2 + |H_\text d|^2) +  W_0 (W_0+ 2 s \mu) (H_\text u H_\text d + \bar H_\text u \bar H_\text d) + \dots \right]\,,
\end{align}
where $T$ is supposed to be stabilized at $t_0$ with a large mass. The ellipsis denotes higher-order terms which are positive definite and unimportant. As before, we neglect the supersymmetric stabilization of $S$ and $U$. We can diagonalize the mass matrix of the fields to find the new states $h = (H_\text u - \bar H_\text d)/\sqrt2$ and $H = (H_\text u + \bar H_\text d)/\sqrt2$. In terms of these the potential reads
\begin{align}\label{eq:v5}
V = \frac{1}{4 s t_0^3}\left[  s^2 \mu^2 |h|^2 + (W_0 + s \mu)^2 |H|^2 + \dots \right]\,,
\end{align}
again neglecting unimportant higher-order interaction terms. Another important ingredient of Higgs-otic inflation is the D-term potential with contributions from both the U(1) charges and the SU(2) charges of $H_\text u$ and $H_\text d$. It was shown in \cite{Ibanez:2014swa} that out of the initial four real neutral scalars one becomes massive due to the D-term potential, while another one corresponding to a Goldstone boson is eaten up by $Z^0$, thereby completing a massive $\mathcal N=1$ vector multiplet. Therefore, only two real scalars remain massless before introducing fluxes, corresponding to  $|h|$ and $|H|$ in the basis of \eqref{eq:v5}. For inflation we thus have to consider the two real degrees of freedom $h \equiv |h|$ and $H \equiv |H|$, or equivalently $\sigma$ and $\theta$ defined by $|H_\text u| = |H_\text d|=\sigma$ and $H_\text u = e^{i \theta}\bar H_\text d$. In terms of the latter the scalar potential can be written as 
\begin{align}
V &= \frac{1}{4s  t_0^3}\left[ s^2 \mu^2 + (W_0 + s \mu)^2 + W_0 (W_0 + 2 s \mu) \cos{\theta}\right] \sigma^2 \nonumber \\
&= \frac{s^2 \mu^2 + (W_0 + s \mu)^2}{4 s t_0^3}\left( 1 + A \cos{\theta}\right) \sigma^2 \,, \label{eq:v6}
\end{align}
where $A \equiv W_0 (W_0 + 2s \mu)/(s^2 \mu^2 + (W_0 + s \mu)^2)$. In general, this potential leads to two-field inflation involving both scalars \cite{Bielleman:2015lka}. The authors of \cite{Ibanez:2014swa} considered two interesting limits:
\begin{enumerate}

\item $A = 0$, corresponding to $GS = 0$ or, equivalently, $W_0 = 0 \lor W_0 = -2 s \mu$. This choice leads to single-field inflation along the direction $\sigma$. $\theta$ is a flat direction in this case.

\item $A =1$, corresponding to $G = S$ or, equivalently, $\mu = 0$. In this case $h$ becomes massless and $H$ is the inflaton with a quadratic potential and $m_H \sim W_0$. Thus, inflation takes place at the same scale as supersymmetry breaking.

\end{enumerate}
We revisit these limiting cases once we have obtained the effective inflaton potential after integrating out the heavy modulus.

\paragraph{Effective inflaton potential} 
$\,$\\
Let us now treat moduli stabilization explicitly. As outlined below \eqref{eq:toy3} we consider the full theory with $H_\text u$, $H_\text d$, and $T$ being dynamical. In the vacuum after inflation $T$ is stabilized by the KKLT mechanism at $t_0$ as discussed in Section \ref{sec:KKLT}, including the uplift to Minkowski space-time. During inflation, as in Section \ref{sec:single}, $T$ couples to both Higgs fields and we must expand $T = t_0 + \delta t(H_\text u, H_\text d)$ and minimize with respect to $\delta t$ to integrate out the modulus consistently. This, again, leads to new terms for both Higgs fields at the level of the scalar potential. After a bit of work we find for the effective potential in terms of $H_\text u$ and $H_\text d$
\begin{align} \label{eq:poteff4}
V &= \frac{1}{8 s t_0^3} \bigg[ (W_0^2 + 2 s \mu W_0 + 2 s^2 \mu^2) (|H_\text u|^2 + |H_\text d|^2) +  W_0 (W_0 - s \mu) (H_\text u H_\text d + \bar H_\text u \bar H_\text d) \nonumber \\
& \hspace{1.2cm} - \mu (W_0 + s \mu)(|H_\text u|^4 + |H_\text d|^4) - \frac32 s \mu W_0  (H_\text u^2 H_\text d^2 + \bar H_\text u^2 \bar H_\text d^2) \nonumber \\
& \hspace{1.2cm} - \frac12 s \mu (5W_0 + 2 s \mu) (|H_\text u|^2 H_\text u H_\text d + |H_\text u|^2 \bar H_\text u \bar H_\text d + |H_\text d|^2 H_\text u H_\text d+|H_\text d|^2 \bar H_\text u \bar H_\text d) \nonumber \\
& \hspace{1.2cm} -5 s \mu(W_0 + s \mu)|H_\text u H_\text d|^2  \bigg] + \dots\,,
\end{align}
once more to leading order in $\alpha t_0$ and $H/m_t$.\footnote{Notice that we same symbol for the Hubble scale and the heavy Higgs field. The respective meaning should be clear from the context.}
This is the two field analog of \eqref{eq:veff}. Notice that most of the quartic terms are now negative and thus are potentially relevant. We can compare the first line to the naive result in \eqref{eq:v4} to see what happened: Due to the re-appearance of a part of $-3|W|^2$ in the effective theory, there is an additional mixed mass term $-3s\mu W_0  (H_\text u H_\text d + \bar H_\text u \bar H_\text d)$ while the other mass terms remain unchanged. The negative quartic terms can similarly traced back to the re-appearance of $-3|W|^2$.

The above expression becomes much simpler when written in the diagonal mass basis. The basis is unchanged from the original Higgs-otic setup discussed above, but the mass eigenvalues are different. We find, instead of \eqref{eq:v5},
\begin{align}\label{eq:v7}
V &= \frac{1}{8 s t_0^3}\bigg[ s \mu (3 W_0 + 2 s \mu) h^2 + (2 W_0^2 + s \mu W_0  + 2 s^2\mu^2) H^2 \nonumber \\
& \hspace{1.2cm} - \frac34 s^2 \mu^2 h^4 - \frac14 s \mu (20 W_0 + 11 s \mu) H^4 + \frac12 s \mu (2 W_0 - s \mu) h^2 H^2 \bigg] +\dots \,.
\end{align}
What we have obtained in \eqref{eq:v7} is, in a way, two copies of \eqref{eq:veff}: Each mass eigenstate has soft mass terms and a dominant quartic term suppressed by one power of $\mu / W_0$. This is exactly the same as in the single-field toy model, and a very intuitive result. It implies that, also in the two-field case, we must require $W_0 \gg \mu$ to guarantee moduli stabilization. Remember that the mass of the modulus is still $m_t \sim W_0$. In the limit $W_0 \gg \mu$, implying $G \approx S$ to high accuracy, two things happen. First, $H$ is stabilized at the origin with a mass $m_H \sim W_0$, at roughly the same scale as the modulus. Second, $h$ is the inflaton and the term proportional to $\mu W_0$ drives inflation---just as in Section \ref{sec:single}. Moreover, this implies that the scale of inflation is suppressed compared to the mass scale of both $T$ and $H$ by one power of $\mu/W_0$, making this a consistent limit. The precise constraint on the superpotential parameters is then 
\begin{align}\label{eq:cons2}
\frac{W_0}{\mu} > \frac{s h_\star^2}{2}\,,
\end{align}
in one-to-one correspondence with the constraint \eqref{eq:constraint2} in the single-field toy model. If it is satisfied, we obtain single-field inflation to very high accuracy. If not, multi-field inflation is nearly impossible with $T$ potentially running away to infinity. Note that again a slightly weaker bound arises from requiring that $\delta t (H_\text u, H_\text d) < 1$, which would be the two-field analog of \eqref{eq:constraint}. Setting $H = 0$ indeed yields the same effective scalar potential for $h$ as in the single-field toy model of Section \ref{sec:single},
\begin{align}\label{eq:veff3}
V_\text{eff}(h) = \frac{1}{4 t_0^3} \left( s \mu^2 h^2 + \frac32 \mu W_0 h^2 - \frac38 s \mu^2 h^4 \right)  + \dots \,.
\end{align}

With these results in mind, let us comment on the two limiting cases of \cite{Ibanez:2014swa}. Due to the backreaction of $T$ the value of the parameter $A$ has slightly changed,
\begin{align}
A=\frac{m_H^2-m_h^2}{m_H^2+m_h^2}=\frac{W_0(W_0-s \mu)}{W_0^2+2 s\mu W_0+2 s^2\mu^2}\ .
\end{align}
This implies the following for the two limiting cases:
\begin{enumerate}

\item $A = 0$ is no longer a consistent option. Both $W_0=0$ and $W_0 = s \mu$ violate \eqref{eq:cons2} when $h>1$ in large-field inflation.

\item While \eqref{eq:cons2} implies that $A$ is very close to one, it is never exactly one. With moduli stabilization taken into account, we must always be in a regime where $h$ drives inflation. If $h$ were to become massless and $H$ were the inflaton, the inflaton and $T$ would have the same mass and $T$ would be immediately destabilized during inflation, as soon as $H \gtrsim 1$. Instead, we may choose $\mu$ very small while $\sqrt{\mu W_0 }$ is the physical mass of the inflaton which is constrained by COBE normalization of the primordial scalar perturbations.

\end{enumerate}
This means that, with this mechanism of moduli stabilization the parameter regime leading to interesting multi-field dynamics---as considered in \cite{Ibanez:2014swa,Bielleman:2015lka}---is excluded.

\begin{figure}[t] 
\centering
\begin{minipage}{0.48\textwidth}
\includegraphics[width=\linewidth]{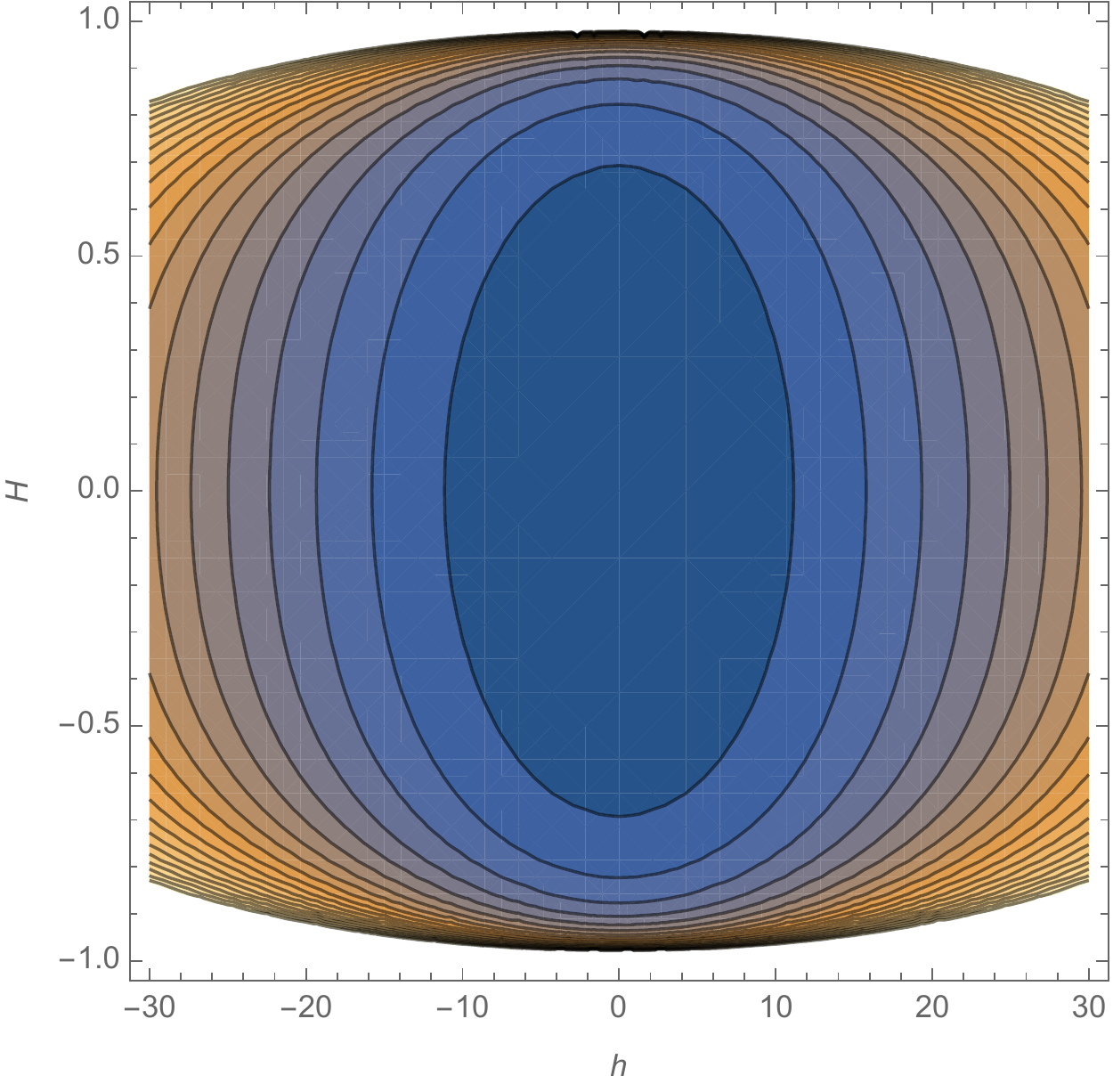}
\end{minipage}
\hspace{0.01\textwidth}
\begin{minipage}{0.48\textwidth}\vspace{-0.4cm}
\vspace{0.5cm}
\includegraphics[width=\linewidth]{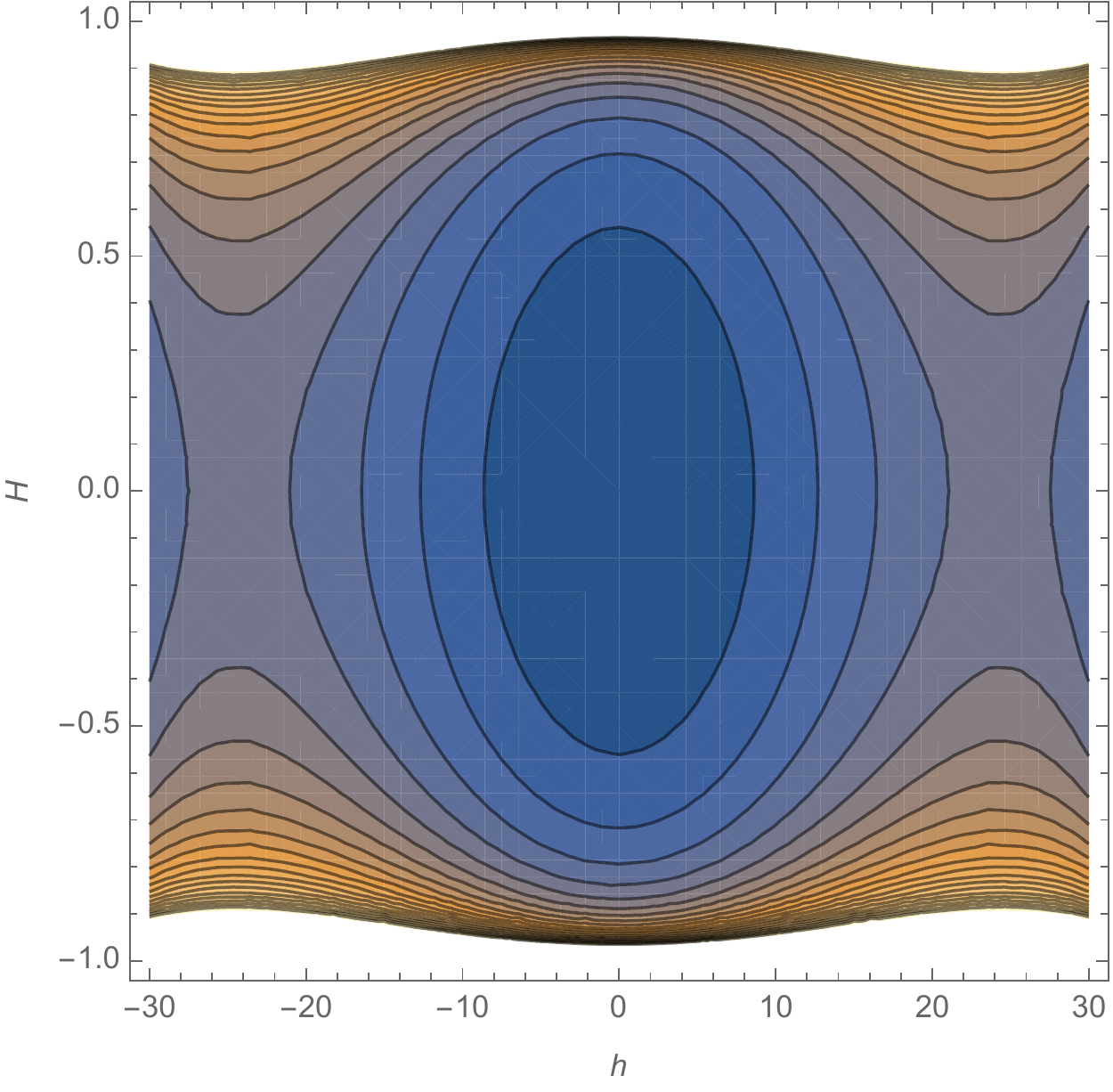}
\end{minipage}
\caption{Contour plot of the original scalar potential $V(h,H)$ from \cite{Ibanez:2014swa} (left panel) compared to effective scalar potential after moduli stabilization (right panel) for the parameter choice \eqref{eq:toypar2}. Warmer color means a larger value of $V$. The darkest blue is the local minimum at $h = H = V = 0$. As expected, the direction $H$ is much steeper than the direction $h$, which is the inflaton direction. In the right panel local maxima are visible at $H=0$ and $|h_\text c| \approx 23$, the point at which the effective theory breaks down and the modulus is destabilized. We have plotted the effective potential \eqref{eq:v7} to all orders in $\alpha t_0$ and $H$, and up to fourth order in $h$. In the case presented here, 60 $e$-folds of slow-roll inflation are possible along the trajectory $H = 0$. The single-field inflaton potential in that slice is identical to the orange line of Figure \ref{fig:veff2}. 
\label{fig:2}}
\end{figure}

\paragraph{A benchmark model} 
$\,$\\
Finally, let us consider a parameter example to illustrate our findings. The $h-H$ plane of the potential is displayed in Figure \ref{fig:2} for the same parameter values as in the single-field example of Section \ref{sec:single},
\begin{align}\label{eq:toypar2}
W_0 = 0.005\,, \quad  \mu = W_0/400\,, \quad s = 1 \,, \quad t_0 = 10\,, \quad \alpha = 2 \pi/5\,.
\end{align}
As expected, the field $H$ has a much steeper potential than $h$ so that $H = 0$ can be a viable inflationary trajectory in this example. In fact the effective theory defined by \eqref{eq:toy3} only describes the potential correctly for $H < 1$ because of a branch cut in the K\"ahler potential, in this case at $H = 1$. To show this behavior, the potential in Figure \ref{fig:veff2} is evaluated to all orders in $H$. In any case, single-field inflation on the trajectory $H = 0$ is a very good approximation. Again, there is a critical field value of the inflaton at which \eqref{eq:cons2} is violated and $T$ is destabilized. It is again $h_\text{c} \approx 23$ for the chosen parameter values. After integrating out $H$ at the origin, $h$ is thus identical to $\varphi$ in our single-field toy model.

%
\subsection{DBI-induced flattening and CMB observables}
\label{sec:DBICMB}

So far we have neglected the nontrivial kinetic term of $\Phi$ in this discussion.\footnote{Here we mean the additional quadratic piece due to the expansion of the DBI action, cf. \eqref{eq:DBICS11}. Additional corrections to the kinetic terms are discussed in Section \ref{sec:CS}.} As discussed in Section \ref{sec:HDO} we can, in principle, include it by adding higher-derivative operators to the supergravity ansatz \eqref{eq:toy1} or \eqref{eq:toy3}. Unfortunately, using the operator \eqref{eq:deltaK} in the single-field case, which corresponds to
\begin{align}
\Delta K = a |\mathbf H_\text u + \bar{\mathbf H}_\text d|^2 \left(  \partial_\mu \mathbf H_\text u \partial^\mu \bar{\mathbf H}_\text u + \partial_\mu \mathbf H_\text d \partial^\mu \bar{\mathbf H}_\text d \right)\,
\end{align}
in the two-field case, does not capture the full result. By stabilizing $T$ non-perturbatively we   break the no-scale symmetry of the effective theory and new couplings involving $T$ and the open-string modulus $\Phi$ appear. We have seen that these couplings modify the effective scalar potential of the inflaton after properly integrating out $T$. We could then expect that the kinetic term of the inflaton is also modified at higher orders in $\alpha'$. However, the higher-derivative operator which captures the correct kinetic term in the $\mathcal N = 1$ supergravity picture should also include the multiplet $T$, not only $\Phi$ or ${\mathbf H}_\text u$ and ${\mathbf H}_\text d$. From the perspective of the worldvolume DBI and CS actions, we have taken modifications into account in Section \ref{sec:IASD}. The gaugino condensate backreacts through the ten-dimensional supergravity equations on the local closed-string background, inducing IASD flux on the bulk. Upon adding this flux in the computation of the effective theory arising from the DBI and CS actions, we found that both the kinetic term and the scalar potential are indeed modified. 

%
\subsubsection{Nontrivial kinetic terms and full Lagrangian}
\label{sec:DBICMB1}

Luckily, due to our analysis in Section \ref{sec:IASD} we do not have to dwell on the intricacies of such complicated higher-derivative theories. Instead, we can read off the correct kinetic term from the flux Lagrangian \eqref{eq:DBICS1} after translating it to our supergravity language. Comparing the potential in \eqref{eq:DBICS12} to the supergravity result after moduli stabilization in \eqref{eq:veff}---similarly for the two-field case in \eqref{eq:poteff4}---leads to the identification
\begin{align}\label{eq:id}
G = \frac{Z W_0}{\sqrt{4 g_\text s s t_0^3}}\,, \quad S = - \frac{Z (W_0 + 2 s \mu)}{\sqrt{4 g_\text s s t_0^3}}\,, \quad D = -\frac{6 Z s \mu}{\sqrt{4 g_\text s s t_0^3}}\,.
\end{align}
Note that the solutions for $G$ and $S$ are the same as the ones in \cite{Ibanez:2014swa} before adding the IASD flux. Inserting the solutions \eqref{eq:id} in \eqref{eq:DBICS1} yields the following leading-order kinetic term for the inflaton field,
\begin{align}\label{eq:kinfin1}
\mathcal L_\text{kin} = -\left(\frac12 + 3a \frac{\mu W_0 \varphi^2}{16 t_0^3} \right)(\partial_\mu \varphi)^2\,,
\end{align}
where $\varphi$ is either the imaginary part of $\Phi$ in the single-field toy model, or is the same as $h$ in the two-field model.\footnote{Note that in the two-field case $H = 0$ always, so that both the potential and the kinetic term of the heavy mass eigenstate are irrelevant during inflation.} Moreover,
\begin{align}\label{eq:a}
a = \frac{1}{V_4 \mu_7 g_\text s} = \frac{16 \pi^3 \alpha_G}{g_\text s M_\text s^4}
\end{align}
as in \cite{Bielleman:2016grv}. In this notation $\alpha_G$ denotes the gauge coupling on the brane stack and ${M_\text s = (\alpha')^{-1/2}}$ is the string scale. Notice the difference between \eqref{eq:kinfin1} and the result prior to moduli stabilization, given by \eqref{eq:kinfinnaive}. The dominant nontrivial piece in the kinetic term is now proportional to $\mu W_0$ instead of $\mu^2$. Thus, it is enhanced by a factor of $W_0/\mu \gg 1$. This is quite intuitive, as the same happened in the scalar potential in our analysis in Section \ref{sec:back}: after taking moduli stabilization into account, the dominant term in the potential is proportional to $\mu W_0$ instead of $\mu^2$. In the kinetic term this can be understood in the following way: the gaugino condensate responsible for stabilizing $T$ sources the additional IASD flux $D$. As we have seen in \eqref{eq:DBICS11}, the kinetic term of the D7-brane modulus then gains additional $D$-dependent terms. Since the gaugino condensate term is proportional to $W_0$ according to the first equality in \eqref{eq:solvac}, the new kinetic term must be proportional to $W_0$ as well. Thus, if we discarded the condensate term and switched off $D$, we would recover the naive result in \eqref{eq:kinfinnaive}---and a destabilized K\"ahler modulus.

In total, the leading-order effective action relevant for inflation is given by
\begin{align}\label{result}
\mathcal L_\text{eff} = -\left(\frac12 + 3a \frac{\mu W_0 \varphi^2}{16 t_0^3} \right)(\partial_\mu \varphi)^2 - \frac{3 \mu W_0 \varphi ^2}{8 t_0^3} + \frac{3 s \mu ^2 \varphi ^4}{32 t_0^3}\,.
\end{align}
Let us now evaluate this result for a few reasonable parameter choices to extract the observables predicted by the combined model.

%
\subsubsection{Parameter examples and CMB observables}
\label{sec:DBICMB2}

\begin{figure}[t] 
\centering
\begin{minipage}{0.48\textwidth}
\includegraphics[width=\linewidth]{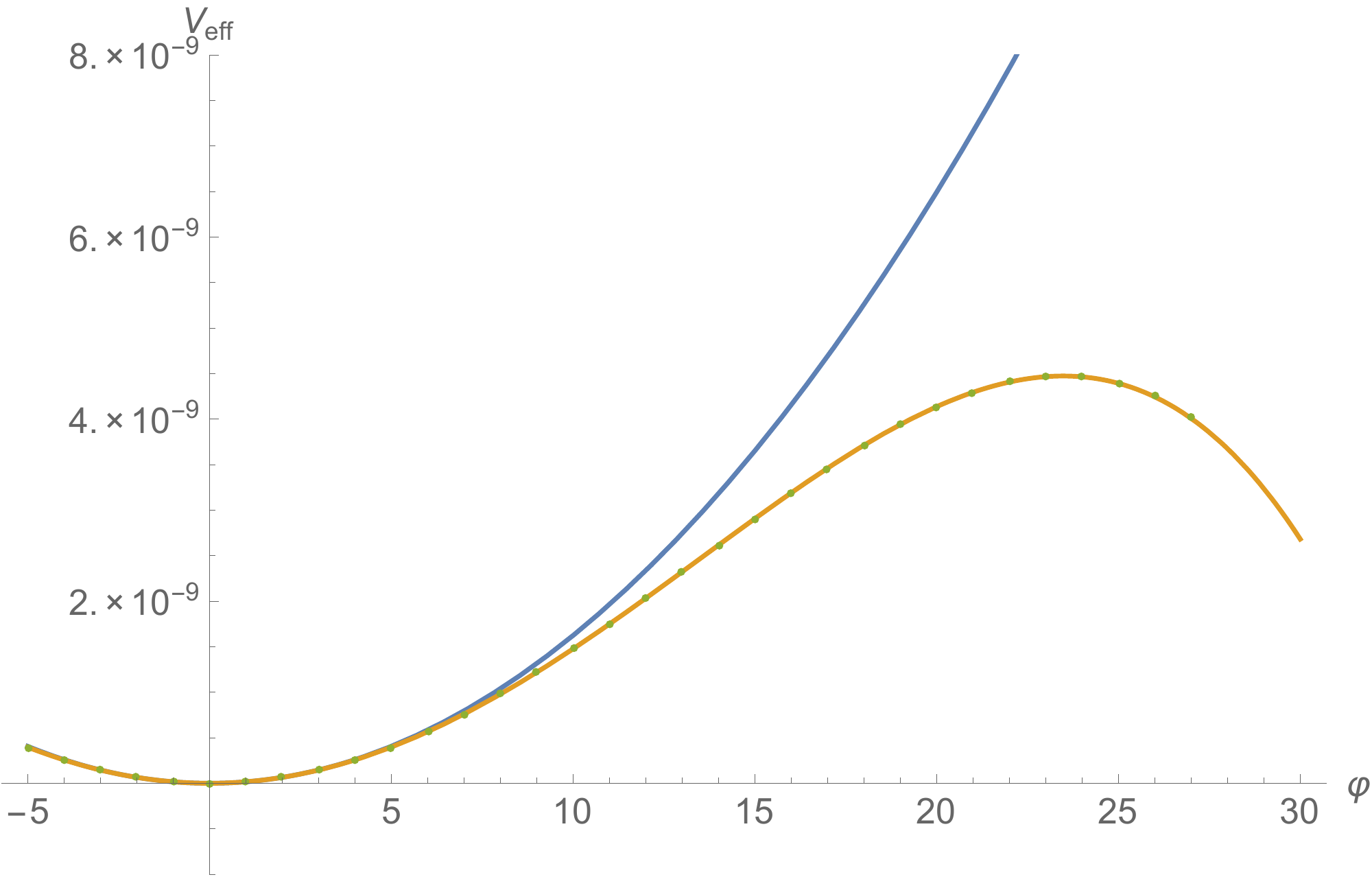}
\caption{Effective scalar potential for the parameter choice \eqref{eq:toypar4}. Naive quadratic potential (blue line) in comparison with effective inflaton potential for $\varphi$ (orange line) and numerical effective potential for the canonical variable in \eqref{result}. The string scale is chosen too large for the DBI-induced flattening to have an effect.
\label{fig:VCMB1}}
\end{minipage}
\hspace{0.01\textwidth}
\begin{minipage}{0.48\textwidth}\vspace{-0.4cm}
\vspace{0.5cm}
\includegraphics[width=\linewidth]{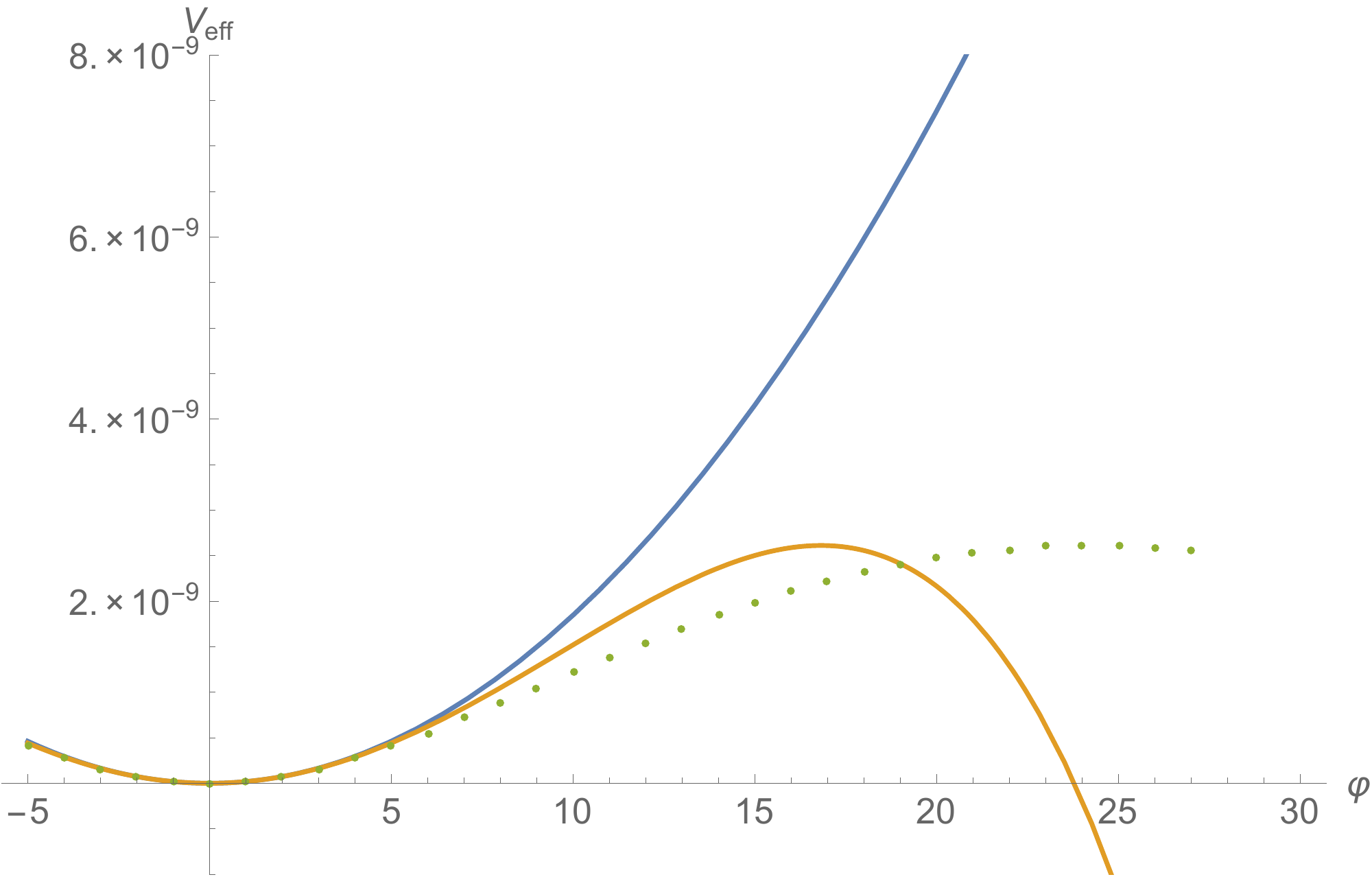}
\caption{Effective scalar potential for the parameter choice \eqref{eq:toypar5}. Naive quadratic potential (blue line) in comparison with effective inflaton potential for $\varphi$ (orange line) and numerical effective potential for the canonical variable in \eqref{result}. In this case the additional flattening from the kinetic term is clearly visible. \label{fig:VCMB2}}
\end{minipage}
\end{figure}

In order to find realistic parameters values, consider the coefficient $a$ in terms of the parameters of the string theory setup in Planck units, \eqref{eq:a}. Let us treat $g_\text s = 0.1$ as a constant and allow ourselves to vary the string scale $M_\text s$ within certain bounds. The most important lower bound is that we must require 
\begin{align}
M_\text s^4 \gg 3 H_\star^2\,,
\end{align}
so that string excitations are negligible during inflation. We must also keep in mind that the compactification scale $M_\text{KK}$ has to fit between the string scale and the inflationary energy scale so that excitations of Kaluza-Klein modes are negligible. In the quadratic approximation we have \cite{Ade:2015lrj}
\begin{align}
V_{\text{inf},\star} = 3 H_\star^2 = 2 \cdot 10^{-11} \cdot 15^2 = 4.5 \cdot 10^{-9}
\end{align}
in Planck units, so that 
\begin{align}
M_\text s \gg V_{\text{inf},\star}^{1/4} = 8.2 \cdot 10^{-3} M_\text p = 1.64 \cdot 10^{16} \,\text{GeV}\,.
\end{align}
To be safe we may choose $M_\text s$ to be larger than $V_{\text{inf},\star}^{1/4}$ by a factor of 10. We thus consider the parameter example 
\begin{align}\label{eq:toypar4}
M_\text s = 0.082\,, \quad \alpha_G = 1/24\,, \quad W_0 = 0.008\,, \quad  \mu = W_0/400\,, \quad s = 1 \,, \quad t_0 = 15\,, \quad \alpha = 2 \pi/5\,,
\end{align}
where the value of $t_0$ is determined by the relation $M_\text s^2 = M_\text p^2 / (8 g_\text s^{1/2} t_0^{3/2})$, cf.~\cite{Ibanez:2012zz}. Note that we have assumed that $\alpha_G$ is independent of $t_0$, i.e., that the gauge theory does not live on the same stack of branes that supports the non-perturbative term in $W$. We can then perform the canonical normalization of the Lagrangian \eqref{result} numerically and plot the resulting potential. The result is given in Figure \ref{fig:VCMB1}. The plot contains the naive quadratic potential (blue line), the effective scalar potential in terms of the variable $\varphi$ after properly integrating out $T$ (orange line), and a numerical plot of the scalar potential in terms of the canonical variable (green dotted line).
Apparently, the flattening induced by the nontrivial kinetic term is a very small effect. The CMB observables on the green dotted line are
\begin{align}
n_\text{s} \approx 0.964 \,, \qquad r \approx 0.087\,,
\end{align}
for 60 $e$-folds of slow-roll inflation at $\varphi_\star \approx 14.6$.

To illustrate the effect the nontrivial kinetic term may have, we can choose a more extreme parameter example: In order to increase $a$ we may increase $\alpha_G$, $s$, and $\mu$ compared to $W_0$, as well as decrease $M_\text s$. Taking the factor between $M_\text s$ and $V_{\text{inf},\star}^{1/4}$ to be 3 instead of 10, we consider the modified parameter set
\begin{align}\label{eq:toypar5}
M_\text s = 0.025\,, \quad \alpha_G = 1/10\,, \quad W_0 = 0.08\,, \quad  \mu = W_0/300\,, \quad s = 2 \,, \quad t_0 = 75\,, \quad \alpha = 2 \pi/5\,.
\end{align}
The relevant potentials can be found in Figure \ref{fig:VCMB2}. As expected, the additional flattening is now much stronger. For the green dotted trajectory one obtains
\begin{align}
n_\text{s} \approx 0.961 \,, \qquad r \approx 0.041\,,
\end{align}
for 60 $e$-folds of slow-roll inflation at $\varphi_\star \approx 12.6$. While this last example illustrates how strongly the kinetic term can effect the CMB observables, it is questionable whether an appropriate hierarchy $M_\text s > M_\text{KK} > V_{\text{inf},\star}^{1/4}$ can be maintained with a value of $M_\text s$ this low. 

%

\section{Complex structure moduli stabilization and backreaction}
\label{sec:CS}

So far we have neglected the stabilization of the complex structure moduli and the dilaton. We have been relying on the expectation that states with masses much larger than the Hubble scale which do not break supersymmetry decouple from the dynamics of inflation \cite{Buchmuller:2014vda}. Recently, however, it was found that in certain type II compactifications backreaction of the complex structure moduli may impair large-field inflation \cite{Baume:2016psm}, see also \cite{Blumenhagen:2015qda,Valenzuela:2016yny}. This can happen whenever the field space metric of the inflaton, in our case
\begin{align}
\varphi = \int \text d \phi \sqrt{ K_{\Phi \bar \Phi}}\,,\label{eq:distance}
\end{align}
with $\varphi$ denoting the canonical distance in field space and $\phi = \text{Im}(\Phi)$, depends on moduli which are displaced during inflation. In our case this applies to $S$ and $U$, but not to $T$. Concretely, since $K_{\Phi \bar \Phi}$ is a function of $S$ and $U$, $K_{\Phi \bar \Phi} \sim (\text{Re}(S) \text{Re}(U))^{-1}$ to leading order, the canonical distance in field space is modified whenever $S$ and $U$ are stabilized such that their expectation value during inflation is $\phi$-dependent. As this is to be expected in most moduli stabilization scenarios, like for $T$  in Section \ref{sec:KMS}, we should treat the stabilization of all moduli explicitly. In \cite{Baume:2016psm} it was argued that in many type II compactifications involving closed-string moduli only, one indeed has schematically
\begin{align}\label{eq:41}
U = u_0 + \delta u (\phi)\,, \quad S = s_0 + \delta s (\phi)\,,
\end{align}
where $\phi$ is a light linear combination of closed-string fields. Moreover, it was shown that beyond a critical inflaton field value where $\delta u (\phi) \gg u_0$ the canonical field distance only increases logarithmically with $\phi$. Depending on the critical value $\phi_\text c$, defined by $\delta u (\phi_\text c) = u_0$, this can make large-field inflation impossible. However, even if the critical field value, and thus the point where the logarithmic dependence dominates, can be tuned large by a flux choice, in the setups of \cite{Baume:2016psm} large-field inflation is under pressure: By tuning $\phi_\text c$ large, $u_0$ and $s_0$ are tuned large as well. Due to the inverse dependence in the K\"ahler metric, this leads to a suppression of the canonical field distance as well. If both $\phi_\text c$ and $s_0,u_0$ parametrically depend on the fluxes in the same way, both effects cancel each other and the canonical field range cannot be larger than the Planck scale.

In this section we demonstrate that these problems are less severe in Higgs-otic inflation. We stabilize $S$ and $U$ via $G_3$ fluxes as in \cite{Giddings:2001yu} and compute the minimum values \eqref{eq:41} explicitly. We show that the fluxes allow for enough freedom to tune $\phi_\text c$ large while, at the same time, leaving $u_0$ and $s_0$ approximately unchanged, so the backreaction effects can be delayed in field space. This flexibility, coming from the introduction of open-string fields, was also discussed in \cite{Valenzuela:2016yny}. Here we aim to make the qualitative arguments given in \cite{Valenzuela:2016yny} more explicit and analyze in detail the resulting back-reacted metric in Higgs-otic inflation. To illustrate these findings we present an analytic study of a useful toy model which is, however, phenomenologically incomplete due to the presence of a flat direction. Afterwards, we turn to a numerical study of a more complete model with full moduli stabilization. 

%
\subsection{A toy model with a flat direction}
\label{sec:toyflat}

Based on a simple toy model taken from \cite{Blumenhagen:2014nba}, we propose the following K\"ahler potential and flux superpotential for a toroidal compactification with a single mobile D7-brane, 
\begin{subequations}
\begin{align}\label{eq:toy2K}
K &= -2 \log(U + \bar U)-\log \left[ (U+ \bar U) (S + \bar S) - \frac12 (\Phi + \bar \Phi)^2 \right] - 3 \log(2 t_0)\,, \\
W &= \mu \Phi^2 + \int G_3 \wedge \Omega = \mu \Phi^2 + e_0 + i m U^3 + i h_0 S + \bar h_0 S U^3\,, \label{eq:toy2W}
\end{align}
\end{subequations}
with integer flux quanta $e_0$, $h_0$, $\bar h_0$, and $m$. Notice the additional term in $K$ compared to \eqref{eq:toy1K}. While this does not change the analysis of Section \ref{sec:KMS}, it accounts for the fact that the compact orientifold in this case is isotropic: we have taken the two-fold $X$ to be a four-torus and have identified the three complex structure moduli of the two-tori. For the moment, we assume that the D7-brane position moduli are stabilized by the presence of (2,1)-fluxes as explained in Section 2, inducing a superpotential term parameterized by $\mu$. Here we treat $\mu$ as an independent parameter and discuss its microscopic origin in terms of NS fluxes in the next section. Furthermore, we consider only the imaginary self-dual piece of $G_3$, so that $D_S W = D_U W = 0$ in the vacuum. On the other hand, the flux potential is non-vanishing in the vacuum, so that supersymmetry is broken and $D_T W \neq 0$. After no-scale cancellation we are thus interested in vacua of the scalar potential
\begin{align}
V = e^K K^{a \bar b}D_a W \overline{D_b W}\,,
\end{align}
where $a$ and $b$ label the fields $\Phi$, $S$, and $U$. We assume that all K\"ahler moduli are stabilized by a KKLT or LVS mechanism as in Section \ref{sec:KMS}. In the remainder of this section we neglect the explicit stabilization and backreaction, which has been analyzed in detail above. It only affects the scalar potential and is irrelevant for the backreaction in the kinetic terms. This is because, on the one hand, \eqref{eq:distance} does not explicitly depend on $T$. On the other hand, the backreaction of $T$ does not affect the backreaction of the complex structure, since the latter only depends on the superpotential and not on the effective scalar potential.

The fluxes in \eqref{eq:toy2W} are only sufficient to stabilize three out of the four real scalar directions. Decomposing $S = s_1 + is_2$ and $U = u_1 + i u_2$ we find the following solutions to the F-term constraints in the vacuum at $\text{Im}(\Phi) \equiv \phi = 0$,
\begin{align}\label{eq:vevs1}
u_{2,0} = 0\,, \quad s_{1,0} = \frac{(e_0 \bar h_0 + h_0 m) u_1^3}{h_0^2 + \bar h_0^2 u_1^6}\,, \quad s_{2,0} = \frac{e_0 h_0 - \bar h_0 m u_1^6}{h_0^2 + \bar h_0^2 u_1^6}\,,
\end{align}
and $u_1$ is a free parameter. During inflation, there is again an interaction between the inflaton $\phi$ and the complex structure moduli. This leads to a modification of the solutions \eqref{eq:vevs1} during inflation,
\begin{align}\label{eq:vevsinf1}
s_{1} = s_{1,0} - \mu \phi^2 \frac{\bar h_0 u_1^3}{2 h_0^2 + 2 \bar h_0^2 u_1^6}\,, \quad s_2 = s_{2,0} - \mu \phi^2 \frac{h_0}{2 h_0^2 + 2 \bar h_0^2 u_1^6}\,,
\end{align}
whereas $u_1$ remains unfixed. These expressions are quite analogous to the displacement of the K\"ahler modulus in  \eqref{eq:deltat}. The inflationary correction is proportional to the Hubble scale, which is determined by $\mu$, divided by the mass of the modulus in question. In particular,
\beq
s_{1} = s_{1,0} - \frac{\bar h_0m_\phi }{8 s_{1,0} m_{s}^2} \phi^2 \label{eq:s1}\,,
\eeq
where $m_s^2 =  ( h_0^2 + \bar h_0^2 u_1^6 ) / (8u_1^3 s_{1,0})$. Thus, by introducing a hierarchy between the mass of the inflaton and the masses of the moduli, we can suppress the displacement compared to the vacuum value of the field. In other words, we can increase the critical value $\phi_\text c$. This can be achieved by tuning $\mu$ to small values compared to $h_0$ and $\bar h_0$. At the same time, this tuning of fluxes does not necessarily affect the vacuum expectation values in \eqref{eq:vevs1} in the same way. $\mu$ does not enter in \eqref{eq:vevs1}, so one can achieve such a mass hierarchy without changing $s_{1,0}$. This is different than in the setups considered in \cite{Baume:2016psm}, where $\phi_\text c$ and $s_{1,0}$ had the same parametric dependence on the fluxes.

Let us now consider the backreaction on the canonical metric in field space during inflation. The metric for the inflaton in field space is given by
\begin{align}
K_{\Phi \bar \Phi} = \frac{1}{4 u_1 s_1}\,,
\end{align}
which indeed decreases for large values of $\phi$ due to the backreaction coming from \eqref{eq:s1}. This implies that, for large field values, the canonical field distance only grows logarithmically with the inflaton field, after performing the integral in \eqref{eq:distance}. However, the point at which the backreaction dominates is flux-dependent, and the canonical field distance travelled before that point,
\beq
\varphi_\text c \approx \frac{\phi_\text c}{4 u_1 s_{1,0}}= \frac{ s_{1,0} m_s }{\sqrt{\bar h_0m_\phi}}\,,
\eeq
can be tuned larger than $M_\text p$ by generating a mass hierarchy between $s_1$ and $\phi$, as we explained above.

In summary, this toy model produces the following result. While the integration of the field space metric does lead to a logarithmic dependence of $\varphi$ on $\phi$ for large field values, the point where the logarithm is relevant can be moved far out in field space by a tuning of fluxes. Of course, the fact that $u_1$ is not stabilized in this setup is a big caveat: we have no means of evaluating the displacement and backreaction of $u_1$ on the possible field space. The above arguments only apply to $s_1$. Furthermore we have not considered the microscopic origin of $\mu$, which could also affect the closed-string fields. This is why, in the following, we use additional fluxes to stabilize $u_1$ and analyze all backreactions simultaneously. A numerical analysis reveals results that are qualitatively the same as in this simple toy model, so the intuition and conclusions reached in this example do not change by considering a more complicated setup.

%
\subsection{Stabilizing all moduli: Flux potential and vacua}
\label{sec:allstab}

To obtain an inflationary theory in which all moduli are stabilized, we must allow for a more general flux $G_3$. We also need to identify the microscopic origin of the $\mu$-term yielding a mass for the D7-brane position moduli. In the Type IIA dual theory involving a wrapped D6-brane one can show that a geometric flux sourcing the term $W \supset a_1 S U $ also induces a supersymmetric mass for the open-string modulus in question. This geometric flux in IIA corresponds to an NS flux on the Type IIB side, and the complexified D6 open-string modulus is mapped to the D7 position modulus in the transverse torus. For more details we refer the reader to Appendix~\ref{app:muterm}.

Let us consider the following effective theory
\begin{subequations}
\begin{align}
K &= -2\log \left[ (U+ \bar U)\right]- \log \left[ (U+ \bar U) (S + \bar S) - \frac12 (\Phi + \bar \Phi)^2 \right] - 3 \log(2 t_0)\,, \\
W &= \mu \Phi^2 + e_0 + i e_1 U + i m U^3 + i h_0 S + \mu S U + \bar h_0 S U^3 \,.\label{W1}
\end{align}
\end{subequations}
where we have also allowed for a linear term in $U$ in addition to the bilinear term $SU$, so that the above superpotential can be written in terms of complexified fluxes pairing $(e_0,h_0)$, $(e_1,\mu)$, and $(m,\bar h_0)$. For the sake of simplicity, we have assumed that the terms $\Phi^2$ and $SU$ have exactly the same coefficient sourced by the same NS flux. However, in more elaborate examples with different complex structure moduli for the three two-tori, $U_1\neq U_2\neq U_3$,  this is not necessarily true. For instance, the position modulus of a tilted D7-brane transverse to one-dimensional cycles of different two-tori feels the presence of NS flux in both two-tori. Thus the coefficient for the $\mu$-term of $\Phi$ would be a combination of several NS fluxes. We come back to this issue when discussing the smallness of $\mu$ in Section 5.

Unfortunately, the two additional terms in $W$ lead to much more complicated equations following from the F-term constraints $D_S W = D_U W = 0$. The solutions can only be studied numerically, which is instructive nonetheless. In particular, the F-term constraints admit a unique Minkowski solution with positive definite Hessian in the vacuum at $\phi = 0$, and at the same time positive vacuum expectation values of the dilaton $s_1$ and the complex structure modulus $u_1$.\footnote{Note that this vacuum is not a deformation of the vacuum found in Section \ref{sec:toyflat}: switching off the two additional flux terms in $W$ does not reproduce \eqref{eq:vevs1}.} 

\begin{figure}[t] 
\centering
\includegraphics[width=0.6\linewidth]{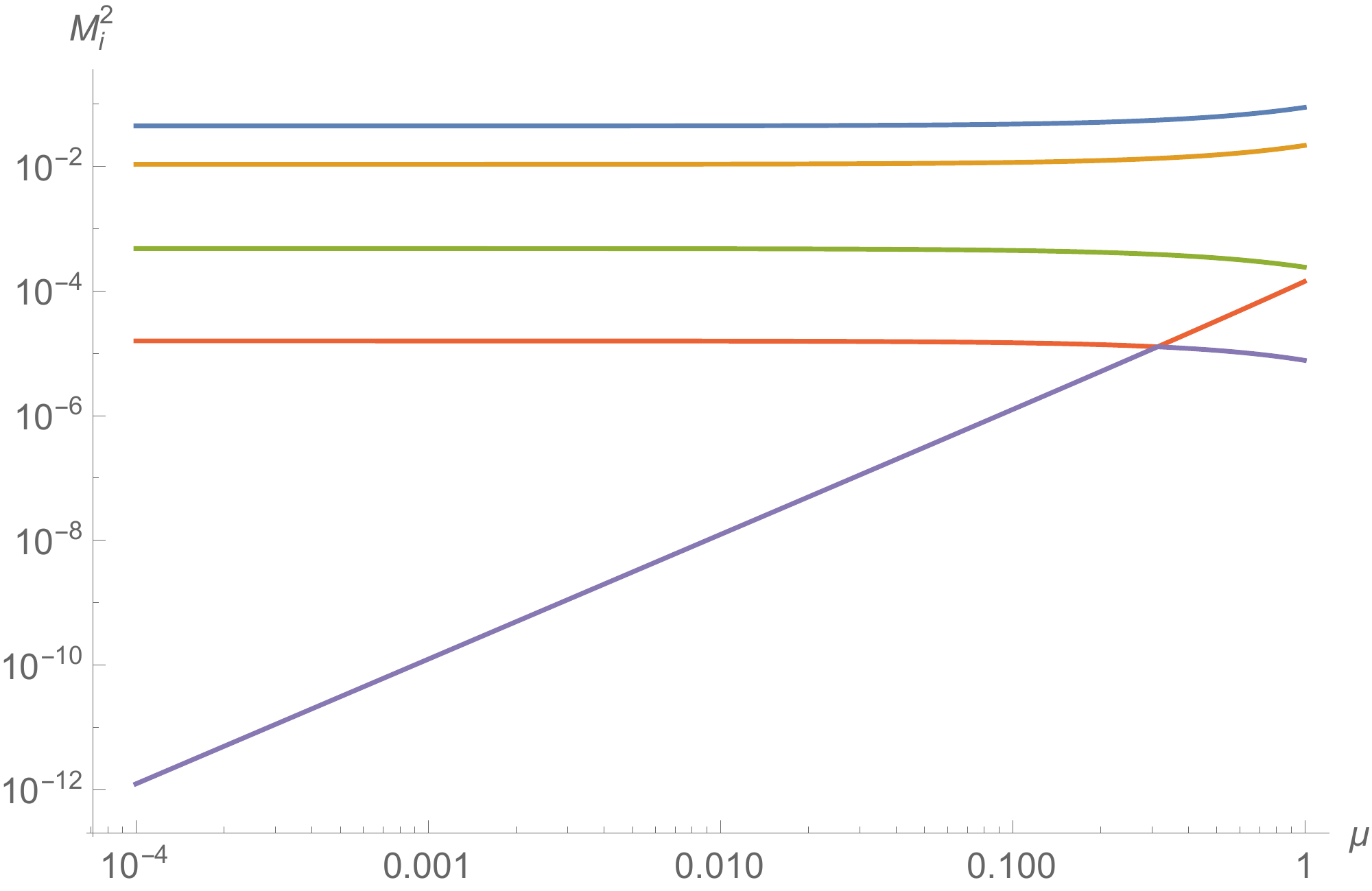}
\caption{Canonically normalized mass eigenvalues of the Lagrangian for the flux choice in \eqref{par1} as well as $t_0 = 30$. The plot is double-logarithmic. The masses are given in units of $M_\text{p}$. Evidently, one mass eigenstate---the would-be inflaton---scales with $\mu$ while the others do not. Making $\mu$ small compared to the other flux parameters introduces a hierarchy between the inflaton and moduli mass scales. \label{fig:EVs1}}
\end{figure}

\begin{figure}[t] 
\centering
\includegraphics[width=0.6\linewidth]{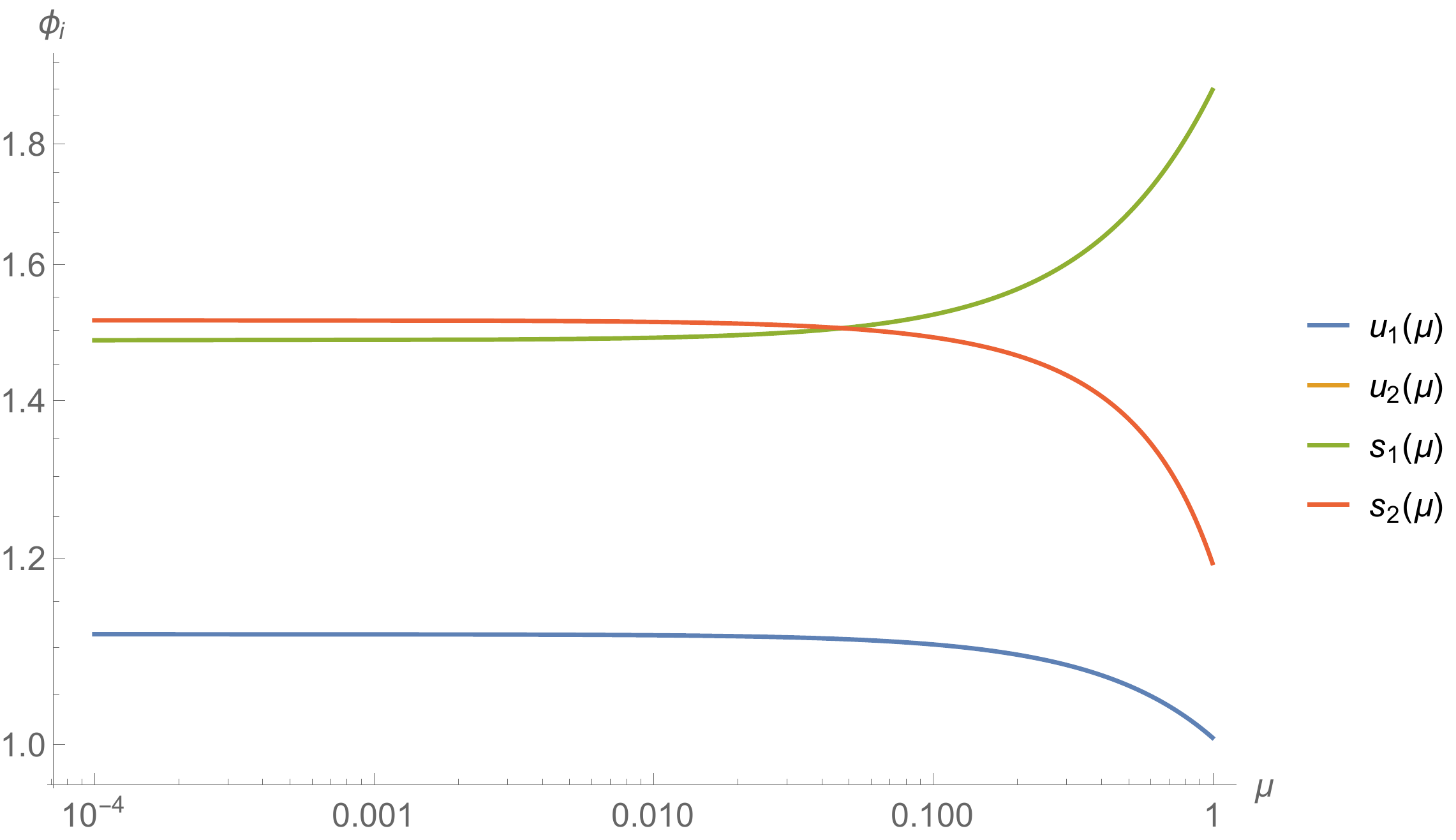}
\caption{Vacuum expectation values of the moduli and axions as a function of $\mu$ with the parameters in \eqref{par1}, as well as $t_0 = 30$ and $\phi = 0$. The plot is double-logarithmic. The value of the axion of $U$ is negative in this example, and therefore is invisible in the logarithmic plot. For small values of $\mu$ the expectation values are independent of $\mu$, contrary to the findings in \cite{Baume:2016psm}. \label{fig:VEVs1}}
\end{figure}

Our goal in the analysis of this unique vacuum solution is to show that there is a hierarchy of masses: When $\mu$ is much smaller than the other flux parameters, one state---the inflaton---is significantly lighter than all others. Moreover, the masses of all four real scalars contained in $S$ and $U$ are mostly independent of $\mu$. To illustrate this, we can plot the five mass eigenvalues of interest ($\text{Re}(\Phi)$ is set to zero and plays no role in this discussion) numerically as a function of $\mu$. We fix all flux parameters except $\mu$ to a set of $\mathcal O(1-10)$ numbers, and vary $\mu$ between $10^{-4}$ and $1$. We choose
\begin{align}\label{par1}
e_0 = -20 \,, \quad e_1 = 20\,, \quad m = 20\,, \quad h_0 = 5 \,, \quad \bar h_0 = -10\,,
\end{align}
as a parameter example. Note that this is equivalent to fixing $\mu$ to an $\mathcal O(1)$ number and varying the remaining parameters to be much larger. What counts is the relative size of $\mu$ compared to the rest of the flux quanta. The result is displayed in Figure \ref{fig:EVs1}. As advertised, the lightest eigenstate has a mass which scales as $\mu$, while the other four eigenstates are much heavier and do not depend on $\mu$ as long as there is a moderate hierarchy between $\mu$ and the remaining flux quanta. This means that, within the flux setup \eqref{W1} we can tune the masses of the moduli and the inflaton independently. Moreover, we can check that the vacuum expectation values of the moduli are almost independent of $\mu$. We have displayed the four vacuum expectation values (at $\phi = 0$) in Figure \ref{fig:VEVs1}, for the same set of flux quanta and the same parameter range of $\mu$.

These results are thus completely analogous to those of the toy model in Section \ref{sec:toyflat}: we can tune the fluxes to obtain a mass hierarchy between the inflaton and the closed-string fields without barely modifying the vacuum expectation values of the latter.

%
\subsection{Backreaction in the K\"ahler metric, flux tuning, and large field excursions}

\begin{figure}[t] 
\centering
\vspace{-0.8cm}
\begin{minipage}{0.48\textwidth}
\includegraphics[width=\linewidth]{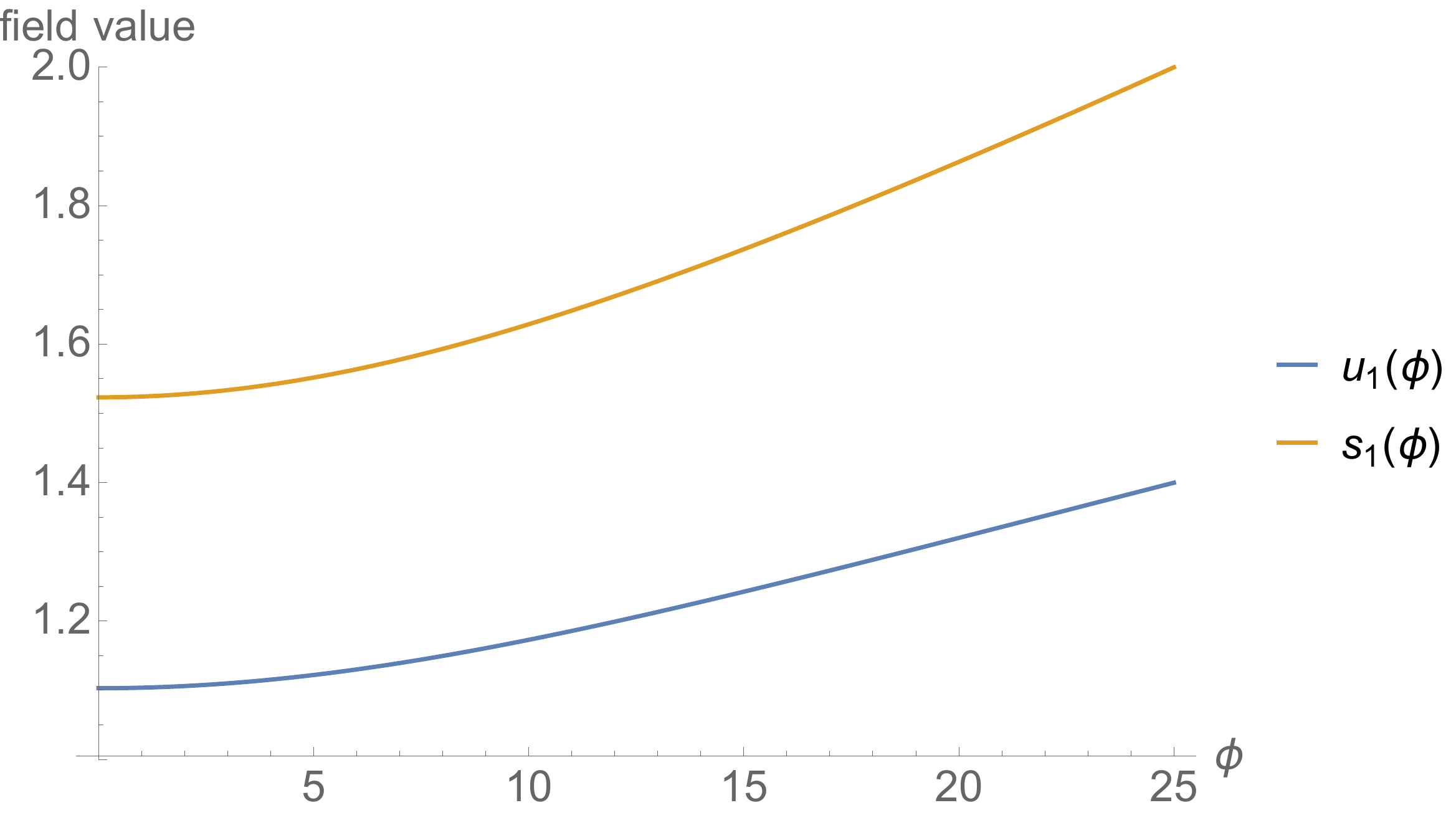}
\caption{Values of the two moduli as a function of $\phi$ for $\mu = 10^{-1}$, $t_0 = 30$, and the flux choice \eqref{par1}.  \label{fig:phi1}}
\end{minipage}
\hspace{0.02\textwidth}
\begin{minipage}{0.48\textwidth}
\vspace{0.5cm}
\includegraphics[width=\linewidth]{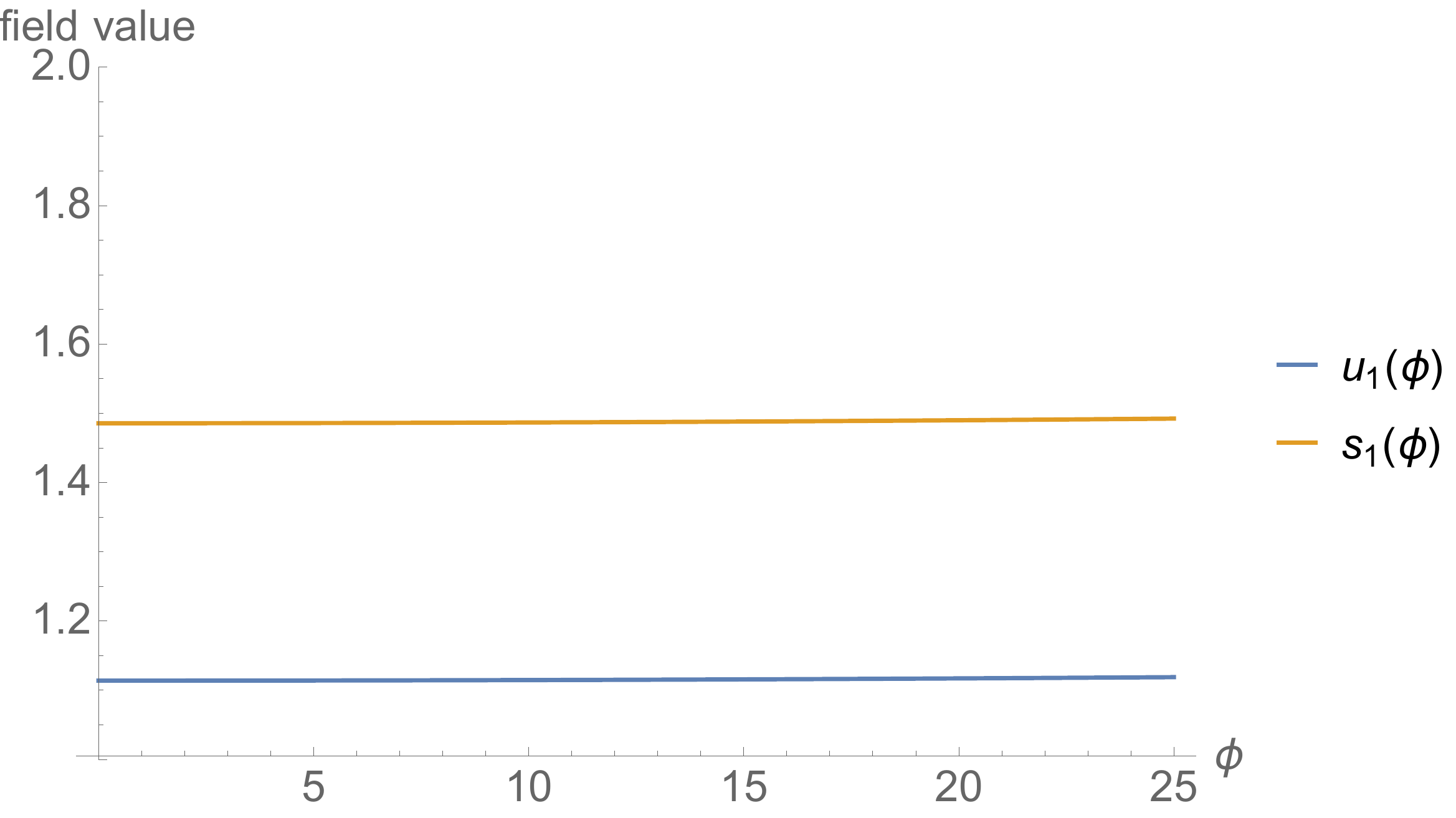}
\caption{Values of the two moduli as a function of $\phi$ for $\mu = 10^{-3}$, $t_0 = 30$, and the flux choice \eqref{par1}. With a smaller value of $\mu$ the two fields are almost constant. \label{fig:phi2}}
\end{minipage}
\end{figure}

Let us now proceed and study what happens during inflation. In Figures \ref{fig:phi1} and \ref{fig:phi2} we have displayed the expectation values of $s_1$ and $u_1$, the only two fields entering the K\"ahler metric of the inflaton, as a function of $\phi$ for two different values of $\mu$. 
Apparently, decreasing $\mu$ (and thus increasing the mass hierarchy) weakens the dependence of $s_1$ and $u_1$ on $\phi$. After what we learned from our toy model in Section \ref{sec:toyflat}, this is no surprise: as in that model---and also in our study of the backreaction of $T$---, increasing the mass hierarchy reduces the field displacements of the rest of the moduli during inflation.

\begin{figure}[t] 
\centering
\includegraphics[width=0.6\linewidth]{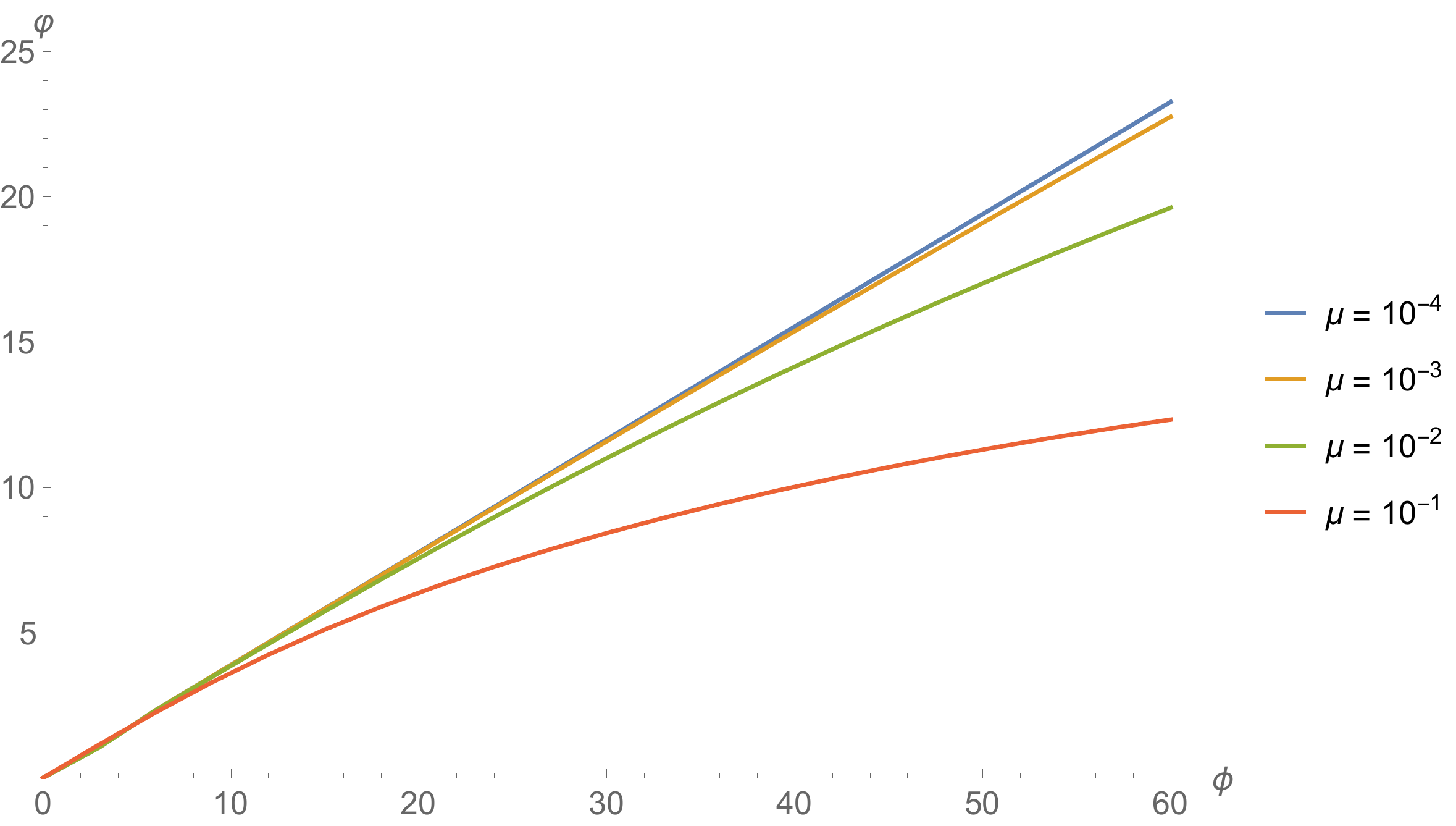}
\caption{Canonically normalized field value as a function of the original variable $\varphi$, for four different values of $\mu$. The logarithmic regime for large values of $\varphi$ is evident in the red and green curves. This is the regime found in \cite{Baume:2016psm}. But evidently, for a large hierarchy between $\mu$ and the remaining flux quanta, the beginning of the logarithmic regime is pushed to very large field values so that the backreaction is negligible. \label{yes}}
\end{figure}

Finally, let us consider what happens to the effective field range
\begin{align}\label{eq:caninf}
\varphi = \int \text d \phi \sqrt{K_{\Phi \bar \Phi}} = \int \text d \phi  \sqrt{\frac{1}{4 s_1(\phi) u_1 (\phi)}}\,.
\end{align}
The important plot is given in Figure \ref{yes}, for four different values of $\mu$. 
Again, the result is quite interesting: For large values of $\mu$, as on the red and green curves, we clearly see the reduced value of $\varphi$ for large $\phi$. As in Section \ref{sec:toyflat} and in \cite{Baume:2016psm}, the $\phi$-dependence of the canonical field distance is logarithmic for large field values. Therefore the backreaction from the closed-string moduli on the field metric of the inflaton makes it difficult to get parametrically large displacements. However, we can push out the canonical critical field value where the logarithmic behavior becomes relevant by decreasing $\mu$ relative to the other flux parameters. This is possible because, unlike in \cite{Baume:2016psm}, the closed-string expectation values $s_{1,0}$ and $u_{1,0}$ do not scale with the fluxes in the same way as the critical value $\phi_c$. Therefore, as long as a tuning of the different flux parameters is possible, the dangerous effect observed in \cite{Baume:2016psm} can be made irrelevant during inflation. 

With regard to inflation, notice that the tuning to reach $\varphi = \varphi_\star \approx 15$ in the nearly-linear regime in Figure \ref{yes} is $\mathcal O(100-1000)$, same as the tuning in the K\"ahler sector in Section \ref{sec:KMS}. In fact, the tuning is not just of the same order of magnitude, it is the same tuning: choosing all flux quanta large compared to $\mu$ leads to a large value of $W_0 = \langle \int G_3 \wedge \Omega \rangle$ compared to $\mu$. In KKLT and related mechanisms, this is exactly what is needed to make $T$ heavy compared to the inflaton.

Finally, one could think of repeating the analysis of the cosmological observables performed in Section \ref{sec:DBICMB} after including the correction to the kinetic term coming from backreaction of $S$ and $U$ in Higgs-otic inflation. However, as long as the aforementioned flux tuning is possible, this correction to the kinetic term is negligible compared to the DBI correction studied in Section \ref{sec:DBICMB}, so the numerical results found there do not substantially change. 

This leads to an interesting alternative to suppress the backreaction in the K\"ahler metric. The $\alpha'$ corrections from the DBI action lead to a nontrivial kinetic term of the form \eqref{result}, which essentially adds an extra term proportional to the DBI scalar potential in the inflaton K\"ahler metric. The resulting effective field range depends on the balance between the DBI correction and the back-reacted saxion expectation values. At the very least, the DBI correction will always help to delay the suppression of the canonical distance and disconnect it from the Planck scale.

%

\section{Discussion\label{sec:discussion}}

The results presented in Sections $\ref{sec:DBI}$--$\ref{sec:CS}$ deserve a careful evaluation. Let us discuss a few aspects in detail:

\begin{itemize}

\item \textbf{Flux tuning}

We have confirmed previous results stating that, in order to guarantee moduli stability during inflation, there needs to be a hierarchy of masses between the inflaton field and all moduli. This applies to the stability of K\"ahler moduli, as discussed in Section \ref{sec:KMS}, as well as to complex structure moduli, as studied in Section \ref{sec:CS}. In the latter we have also checked that this mass hierarchy is enough to suppress the backreaction on the field-space metric and the canonical field distance. In our particular model of Higgs-otic inflation, this hierarchy must be achieved by a tuning of fluxes: in one case we tune $\mu$ small compared to $W_0$, in the other case we tune $\mu$ small compared to all other flux parameters entering the Gukov-Vafa-Witten superpotential. Therefore, in both cases one may ask how $\mu$ can be a number as small as $10^{-4}$ or $10^{-5}$.

As is well known, fluxes are determined by integrals of $p$-form gauge fields over internal $p$-cycles and, therefore, must obey a Dirac quantization condition, i.e., $\mu_{p-2}\int_{\Sigma_p} G_p=2\pi n$ with $n\in \mathbb{Z}$. Here $\mu_p=(2\pi)^p\alpha'^{(p+1)/2}$ is the charge of the corresponding object electrically charged under the gauge field. This implies that all flux quanta $e_0$, $e_i$, $m$, $h_0$, and $\bar h_0$ are indeed integers. If $\mu$ is also a flux quantum, a value of $10^{-4}$ is not allowed. However, the $\mu$-term for the open-string modulus in the superpotential can receive contributions from different sources and fluxes in the compactification. By requiring some fine-tuning between the different contributions one can obtain smaller values of $\mu$. This is analogous to the rationale behind the KKLT mechanism and the fine-tuning arguments to obtain a small value of $W_0$, cf.~the discussions in \cite{Bousso:2000xa,Denef:2004ze} and also \cite{Cicoli:2013swa}. For instance, in the case of a tilted D7-brane in a toroidal compactification, the position modulus feels the presence of NS flux in the two two-tori that are only partially wrapped by the brane. Another possibility is to consider compactifications beyond toroidal models, with a large number of three-cycles leading to many contributions to the $\mu$-term. From the perspective of the brane worldvolume action, the brane is only sensitive to the local background and the flux densities around the brane. These are a combination of many different internal fluxes as well as distant sources back-reacting on the local background. Therefore, a priori there is no restriction to a small value of $\mu$. However, increasing the number of three-cycles implies a larger number of complex structure moduli. Even if this does not change the results in Section \ref{sec:KMS}, it might make the analysis in Section \ref{sec:CS} intractable. Nevertheless, the leading source of mass terms for the complex structure moduli can---as in our example above---come from a set of fluxes which do not affect the D7 position moduli and thus do not contribute to the $\mu$-term. Then the expectation values of the complex structure moduli are still approximately independent of the value of $\mu$. In that case, we expect the conclusions of Section \ref{sec:CS} to be qualitatively unchanged in more complicated models. 

Another possibility would be not to decrease $\mu$, but to increase the value of the other fluxes. Not the absolute values but the ratios to $\mu$ are what matters in suppressing the backreaction. However, large fluxes are problematic for several reasons. They can easily lead to moduli masses heavier than the KK scale and, furthermore, yield a big RR tadpole which has to be cancelled by a large number of sources with negative D3-brane charge. The backreaction of these fluxes and additional sources in the global compactification might force us to work beyond the validity of our effective theory.

\item \textbf{Magnitude of the flux superpotential}

There is another question closely related to the previous issue: can the benchmark models we have presented in Sections \ref{sec:KMS} and \ref{sec:CS} be made compatible? 

The answer seems to be nontrivial, since our example in Section \ref{sec:CS} implies---mostly independent of the value of $\mu$---a large expectation value of the flux superpotential, 
\begin{align}
|W_0| =\left| \langle \int G_3 \wedge \Omega \rangle \right| \approx 60\,,
\end{align} 
in Planck units. The KKLT mechanism used in Section \ref{sec:KMS}, on the other hand, requires $W_0$ to be small compared to $M_\text p^3$. There are two ways out of this predicament. First, in a more complicated compactification with more flux parameters, $W_0$ may be small even though all flux parameters are integer. This is very similar to the way $\mu$ can be made small, as discussed above. Thus, there may be a scenario in which both $W_0 \ll 1$ and $\mu \ll 1$ while the mass hierarchy between inflaton and complex structure moduli is unchanged. In this case $T$ can be stabilized by the KKLT mechanism as in Section \ref{sec:KMS}, and the inflaton potential $V \sim \mu W_0 \varphi^2$ may have the correct normalization.

Second, even if $W_0$ is $\mathcal O(1-10)$, there are viable mechanism for K\"ahler moduli stabilization available. Both the Large Volume Scenario \cite{Balasubramanian:2005zx,Conlon:2005ki} and K\"ahler Uplifting \cite{Balasubramanian:2004uy,Westphal:2006tn} do not require a tuning of $W_0$. In fact, for both mechanisms the interaction with open-string large-field inflation has been studied in \cite{Buchmuller:2015oma}. The results are very similar to the KKLT scenario: inflation is mostly driven by an inflaton mass term proportional to $\mu W_0 \varphi^2$, and $W_0 \gg \mu$ is required to guarantee stability of all K\"ahler moduli in the inflationary phase. Also in this case there is a certain range of parameter examples that lead to 60 $e$-folds of slow-roll inflation in accordance with CMB observations.

Thus, the two tunings in Section \ref{sec:KMS} and \ref{sec:CS} are indeed compatible, so that all moduli can be stabilized during Higgs-otic inflation.

\item \textbf{Constraints on the canonical field space}

Let us discuss another issue related to the findings of Section \ref{sec:CS}: why does Higgs-otic inflation not exhibit the restriction of the canonical field space found in \cite{Baume:2016psm}? As explained in the very beginning of Section \ref{sec:CS}, the problems in \cite{Baume:2016psm} are not only related to mass hierarchies. As nicely pointing out in that paper, and more recently in \cite{Valenzuela:2016yny}, even if such a mass hierarchy can be achieved, one also has to check that the post-inflationary vacuum expectation values of some of the moduli entering the K\"ahler metric of the inflaton do not scale with this hierarchy as well. In our example this would correspond to an increase of $s_{1,0}$ and $u_{1,0}$ as the mass hierarchy is enhanced. Such an increase would then lead to a reduction of the canonical field range due to \eqref{eq:caninf}. In the setups of \cite{Baume:2016psm}, this reduction indeed cancels the increase obtained by engineering a mass hierarchy. Therefore the canonical field range before the backreaction becomes strong is flux-independent and cannot be larger than the Planck scale. However, as we have demonstrated both in the toy model of Section \ref{sec:toyflat} and the numerical example of Section \ref{sec:allstab}, this does not happen in Higgs-otic inflation as long as a hierarchy of fluxes is present: there are enough parameters to treat the mass hierarchy and the vacuum expectation values of the relevant fields as independent parameters.

The reason for this lies in the respective microscopic setups: the ones considered in \cite{Baume:2016psm} feature closed-string moduli only. The inflaton candidate is then a light linear combination of otherwise heavy axions. This, in turn, leads to no-go results for large-field inflation similar to those found in \cite{Blumenhagen:2014nba,Hebecker:2014kva}. In contrast, Higgs-otic inflation features a completely independent light degree of freedom: the position modulus of the D7-brane. Tuning the parameter $\mu$ compared to the other flux parameters---which determine the masses and vevs of the closed-string fields---corresponds to a freedom that is not present in the analysis of \cite{Baume:2016psm}. We should remark, though, that the introduction of open-string fields also makes the analysis of the global compactification more difficult, including the effect of all possible backreactions. Still, we find that the canonical field range indeed scales logarithmically with the inflaton field value in the large-field limit. But the physical critical value at which this logarithmic behaviour becomes relevant is flux-dependent and therefore not necessarily tied to the Planck scale.

\item \textbf{Matching the Higgs mass}

The results obtained in Sections \ref{sec:KMS} and \ref{sec:CS} not only apply to Higgs-otic inflation, but to any inflationary model in which the inflaton is identified with a D7-brane position modulus in a similar background. In fact, in order to keep the discussion as generic as possible, we have only imposed constraints from cosmological data and not from particle physics so far. Therefore, let us discuss now if the results obtained in the previous sections are still compatible with particle physics phenomenology, if we identify the inflaton with the SM Higgs boson as in \cite{Ibanez:2014swa}. 

In order to keep one mass eigenstate---the SM Higgs boson---light at low energies, there must be an almost massless field at the supersymmetry breaking scale $M_\text{SS}$, below which the supersymmetric spectrum decouples. This happens when the running of the soft mass parameters from the compactification scale $M_\text c$ down to $M_\text{SS}$ gives rise to a zero eigenvalue in the Higgs mass matrix, i.e., $\text{det}(M_H^2)=m^2_{H_\text u}m^2_{H_\text d}-m_3^4\approx 0$ at $M_\text{SS}$. For a given value of $M_\text{SS}$ this imposes a constraint on the mass ratio $A=m_3^2/m_{H_\text u}^2$ at $M_\text c$. It was shown in \cite{Ibanez:2013gf} that, if such a fine-tuned Higgs survives, one necessarily gets $m_h=126\pm 3$ GeV for $M_\text{SS}=10^9-10^{13}$ GeV and standard unification conditions $m_{H_\text u}=m_{H_\text d}$ at $M_\text c$. The question is whether this constraint on the mass ratio $A$ is compatible with the mass hierarchy required to get moduli stability during inflation.

Specifically, from \eqref{eq:v7} we can derive the value of $A$ in terms of $\mu$ and $W_0$, 
\beq
A=\frac{m_3^2}{m_{H_\text u}^2}=\frac{m_H^2-m_h^2}{m_H^2+m_h^2}=\frac{W_0(W_0-s \mu)}{W_0^2+2 s\mu W_0+2 s^2\mu^2}\,.
\eeq
The two benchmark points considered in Section \ref{sec:KMS} correspond to $A=0.993$ with $M_\text{SS} \approx 9\cdot 10^{13}$ GeV and $A=0.990$ with $M_\text{SS} \approx 5\cdot 10^{13}$ GeV, respectively. As already mentioned in Section \ref{sec:KMS}, the mass hierarchy required to get moduli stability and suppress the backreaction of $T$ implies a value of $A$ very close to one. This, in turn corresponds to an almost massless state already at $M_\text c$. Unfortunately, the above values of $A$ are too large and very little running is required to make the Higgs determinant vanish. Therefore, the massless eigenstate will appear close to $M_\text c\approx  10^{16}$ GeV, implying that the Higgs boson at $M_\text{SS}$ is already tachyonic and triggers electroweak symmetry breaking at a too high energy scale.

Let us remark that we have assumed no additional physics until $M_\text{SS}$, and only the MSSM spectrum beyond it. Additional states at high energies could modify the renormalization group equations for the soft mass parameters, leading to less stringent constraints on the value of $A$. Furthermore, the above tension arises from the fact that the mass scale for the heavy Higgs $H$ coincides with the mass scale of the K\"ahler modulus, parameterized by $W_0$. If one finds a scenario where both scales are decoupled, one could decrease the mass of $H$ while maintaining $\mu/W_0\ll 1$ and moduli stability. That, in turn, would lead to a lower value of $A$. This might be possible, for instance, by placing the system of D7-branes in a strongly warped region of the compactification. 

\end{itemize}
%

\section{Conclusions}

We have studied moduli stabilization and its effects in large-field D7-brane inflation in Type IIB string theory. We have focussed on the setup of Higgs-otic inflation, in which the position modulus of one or more D7-branes is a monodromic axion with potentially trans-Planckian field excursion. The monodromy is sourced by three-form fluxes in the internal manifold. The most important conclusion of our work is that, provided a certain tuning of fluxes is possible, large-field inflation and full moduli stabilization can be successfully combined. 

We have analyzed the stabilization of K\"ahler moduli via non-perturbative physics, and---confirming earlier results in the literature---have shown that stability is guaranteed throughout the expansion history of the universe as long as the mass of the moduli can be tuned larger than the inflationary Hubble scale at horizon crossing. In the particular setup at hand this means that the flux density $\mu$, responsible for the mass of the inflaton, must be tuned much smaller than the flux superpotential $W_0$, which determines the mass of the volume modulus. For an appropriate hierarchy between these two scales, in our examples a factor 300 or 400, integrating out the heavy modulus in its inflaton-dependent minimum leads to a correction of the inflaton potential. This correction is negative and leads to a flattening of the potential. This effect has been observed before in the literature, and seems generic in large-field inflation models from string compactifications.

In addition, the inflaton has a nontrivial kinetic term because its dynamics is governed by the DBI action. We have shown that gaugino condensates responsible for moduli stabilization lead to a coefficient of the kinetic term that is proportional to the leading-order scalar potential, after the backreaction of the volume modulus is taken into account. This coefficient leads to an additional flattening of the potential. Through a matching of the flux parameters in the expanded DBI action with the supergravity data, we have numerically obtained the action for the canonically normalized inflaton field. For large field values, the shape of the effective potential approximates $V(\varphi) = a \varphi - b \varphi^2$. The predictions for 60 $e$-folds of slow-roll inflation lie well within the allowed regime of the latest CMB data.

Moreover, we have discussed the stabilization and backreaction of complex structure moduli and the dilaton. We have analyzed scenarios in which both can be stabilized supersymmetrically by additional three-form fluxes. Regarding stability and backreaction of the moduli, the results are quite similar to the K\"ahler sector. If the mass hierarchy between the inflaton and the moduli can be made sufficiently large---again $\mathcal O(100-1000)$ in this case---all backreaction effects are negligible. In particular, this applies to a specific backreaction in the kinetic term which, in scenarios with only closed-string fields, leads to a severe reduction of the canonical field distance traveled during inflation. As we have demonstrated, in our setup this reduction is negligible during large-field inflation. The inclusion of open-string fields generically yields more freedom to tune the necessary hierarchies.

%

\section*{Acknowledgments}

We thank David Ciupke, Aitor Landete, Fernando Marchesano, and Eran Palti for useful discussions. This work is partially supported by the grants FPA2012-32828 from the MINECO, the ERC Advanced Grant SPLE under contract ERC-2012-ADG-20120216-320421 and the grant SEV-2012-0249 of the ``Centro de Excelencia Severo Ochoa" Programme. I.V. is supported by a grant from the Max Planck Society.

%
\appendix


%
\section{Fluxes and $\mu$-terms}
\label{app:muterm}

We provide here a toroidal Type IIB orientifold example showing how certain closed-string fluxes not only contribute to the moduli superpotential but also generate $\mu$-terms for charged matter fields. To be concrete, consider a toroidal setting $T^2\times T^2\times T^2$ with the standard $O(3)$ orientifold projection. Then consider NS fluxes,
\beq
{H}_3  =   -\sum_{i=1}^3 a_i\a_i\,,
\label{hfluxb}
\eeq
which are expanded in the standard basis of three-forms on the torus, following the notation of \cite{Ibanez:2012zz}. Then the Gukov-Vafa-Witten superpotential yields
\beqa
W  =  -\sum_i a_iSU_i  \, ,
\label{wnsrr}
\eeqa
where $S$ is the complex dilaton and $U_i$ are the complex structure moduli of the three tori.  Consider now a D7-brane wrapping the first two tori and transverse to the third. We want to show that the same NS fluxes induce a $\mu$-term for the adjoint position modulus $\Phi_3$ which parameterizes the position of this D7-brane in the transverse torus.  

For convenience we  start with the mirror Type IIA toroidal orientifold with an $O(6)$ projection and a D6-brane wrapping, for example, the cycle
\beq 
\Pi_3  = (0,1)_1\times (0,-1)_2\times (1,0)_3\,,
\eeq
where $(n,m)$ means  that the D6-brane wraps $n$-times around the $x$ direction and $m$ times around $y$. In the IIA mirror the NS  fluxes above map into geometric fluxes. For simplicity, let us consider only the flux $a_3$. The mirror IIA geometric flux is $\omega_{45}^3= a_3$, in the notation of \cite{Ibanez:2012zz}. Let us consider now the Chern-Simons coupling on the worldvolume of the D6-brane,
\beq
\int_{\Pi_3\times M_4} C_3\wedge F\wedge F   \, .
\eeq
In the presence of geometric fluxes $\omega_{ab}^i$ or, equivalently, on a twisted torus one replaces
\beq
F _{ab} \to  F_{ab} + \omega _{ab}^iA_i  =F_{ab}  + (\omega . A)_{ab}\,,
\eeq
so that, after putting the legs of $C_3$ in the Minkowski direction and integrating by parts, one gets
\beq
F_4^0 \int_{\Pi_3} ( \omega  .A) \wedge A\,,
\eeq
where $F_4^0$ is a Type IIA Minkowski four-form. We thus see that there is a coupling of this four-form to a Wilson line bilinear controlled by the background $\omega$. In particular, for the three-cycle above one finds the action
\beq
a_3 F_4^0 \text{Tr}(A_3)^2 = a_3 F_4^0 \text{Tr}(\theta _3)^2 \, ,
\eeq
where $\theta_3$ is a Wilson line scalar on the D6-brane. As discussed in \cite{Bielleman:2015ina}, in Type IIA $F_4^0$ couples to the real part of the superpotential, i.e.
\beq
F_4^0 \, \text{Re}(W)\,,
\eeq
so we can identify a contribution to the superpotential
\beq
\text{Re}(W_{a_3}) = a_3 \text{Tr}(\theta _3)^2 \,.
\eeq
Holomorphicity allows us to complete the form of the induced superpotential. Along the third torus the 
D6-brane open-string modulus is a combination of the Wilson line $\theta_3$ and position modulus $\phi_3$, which parameterizes the motion in the direction perpendicular to the D6-brane in that complex plane:
\beq 
\Phi_3 = \theta_3  + T_3\phi_3\,.
\eeq
Hence, the piece of the superpotential proportional to $a_3$ is
\beq
W_{a_3} = a_3 \Phi_3^2\,,
\eeq
which is a $\mu$-term.  Let us check for completeness that the cross term in $\Phi_3^2$ involving the coupling $\theta_3 \text{Re}(T_3) \phi_3$ also appears in the action. In addition to the above CS coupling, there is a coupling on the twisted torus of the form
\beq
\int_{\Pi_3\times M_4} C_3\wedge  F \wedge [\omega .B]_P\,,
\eeq
where the subscript $P$ indicates the pullback and $B$ is the NS two-form. After partial integration we get a coupling for the above choice of D6-brane,
\beq
F_4^0  \int_{\Pi_3} A\wedge [\omega .B] \phi_3  =  F_4^0 a_3 \text{Tr}(\theta_3 b_3 \phi_3)\,,
\eeq
where $T_3=b_3+ iJ_3$ and $\phi_3$ is the position modulus transverse to the D6-brane in the third complex plane. We observe that the required term is in indeed present.

Going back to IIB, dualizing along the three horizontal directions of the torus, we end up with a D7-brane which is localized on the third torus and wraps the other two. The field $\Phi_3$ is now mapped to a complex scalar which parameterizes the position on the third torus. We conclude that in IIB the standard NS flux $a_3$ gives rise not only to a moduli superpotential piece $W \sim a_3 SU_3$ but also to a contribution to the $\mu$-term for the adjoint $\Phi_3$.


\end{document}